\documentclass[aps,prb, reprint,superscriptaddress,longbibliography]{revtex4-2}
\usepackage{amsmath,amssymb,bm}
\usepackage{graphicx}
\usepackage{dcolumn}
\usepackage{hyperref}
\usepackage{xcolor}
\usepackage{physics}
\usepackage{braket}
\usepackage{mathtools}
\usepackage{float}
\usepackage{booktabs}

\usepackage{hyperref}
\hypersetup{
colorlinks=true,
linkcolor=blue!60!blue,
citecolor=blue!60!blue,
urlcolor=blue!60!blue}

\usepackage[skip=2pt]{caption}
\usepackage[justification=raggedright,singlelinecheck=false]{caption}

\begin{document}

\title{Topological signatures in the quench dynamics of periodically driven quantum systems}

\author{Ashutosh Dubey}
\affiliation{Department of Physics, Indian Institute of Science Education and Research Tirupati, Tirupati 517619, India}
\affiliation{Department of Physics, Indian Institute of Technology Kanpur, Kanpur 208016, India}
\author{Hironmoy Pratihar}
\affiliation{Department of Physics, Indian Institute of Technology Kanpur, Kanpur 208016, India}
\author{Diptiman Sen}
\affiliation{Centre for High Energy Physics, Indian Institute of Science, Bengaluru 560012, India}
\author{Arijit Kundu}
\affiliation{Department of Physics, Indian Institute of Technology Kanpur, Kanpur 208016, India}

\begin{abstract}
We study the quench dynamics of graphene, without and with a staggered mass, following the sudden switch‑on of circularly polarized light where coupling to a fermionic bath is also considered. Using an armchair nanoribbon, we compute the period‑averaged bond current near the edge at successive stroboscopic times. For the isolated system, the current oscillates around a dc value which, we analytically show, equals the Floquet band currents weighted by their projected occupations; a nonzero dc current signals an induced topological phase. When coupled to a bath with a finite coupling, the current and conductance initially increase and then saturate, indicating a nonequilibrium steady state. In the limit of vanishingly small coupling, the period‑averaged conductance becomes quantized after summing over bath chemical potentials shifted by integer multiples of the driving frequency, revealing the number of edge modes crossing the zero‑quasienergy gap and the Floquet zone boundary gap. We support these results by computing the two‑terminal conductance of a finite‑size tight‑binding model under a three‑step driving protocol; the bond conductance after applying the sum rule corroborates our findings.
\end{abstract}

\maketitle

\section{Introduction}

Understanding quantum phases of matter has been a central theme in condensed matter physics research. In particular, topological phases~\cite{RevModPhys.89.041004, PhysRevLett.95.226801, PhysRevLett.98.106803, PhysRevB.83.205101, kitaev2001unpaired}, which lie beyond the conventional Landau paradigm of spontaneous symmetry breaking, have attracted significant attention due to their fundamental importance and potential applications in the field of quantum information~\cite{Lahtinen2017, PhysRevB.87.195451} and quantum computation~\cite{10.21468/SciPostPhys.3.3.021}. Considerable efforts have been devoted to realizing, manipulating, and controlling such phases in both real materials~\cite{C3TC30186A} and quantum simulators~\cite{PhysRevLett.122.210503, Pickup2020, ArguelloLuengo2024}. Among the various approaches, quantum quenches~\cite{annurev:/content/journals/10.1146/annurev-conmatphys-031016-025451} and periodic driving have emerged as powerful tools for engineering and controlling quantum systems, enabling the realization of novel phases and dynamical phenomena that are inaccessible in equilibrium settings.

Periodic driving has emerged as a powerful tool for manipulating and controlling quantum systems. In such setups, a quantum system is subjected to an external time-periodic perturbation, such as circularly polarized light, and its dynamics are commonly described within the framework of Floquet theory~\cite{eckardt2015high, Bukov2015,sen2021}. A periodic drive can significantly modify the band structure by renormalizing hopping amplitudes and opening gaps in the band structure~\cite{PhysRevB.79.081406, PhysRevLett.113.236803, Rudner2020, PhysRevResearch.4.033213, PhysRevResearch.2.043275}. These effects have been directly observed in angle-resolved photoemission spectroscopy (ARPES) experiments~\cite{doi:10.1126/science.1239834, RevModPhys.96.015003}. As a consequence of this Floquet band engineering, periodic driving can induce a variety of nonequilibrium phases, including Floquet topological insulator~\cite{Lindner2011, PhysRevB.92.165111}, Floquet Weyl semimetal~\cite{Hubener2017}, and Floquet superconductor~\cite{Kitamura2022}, and many exotic phases~\cite{doi:10.1126/sciadv.aay4922, Esin2021, Fausti2011, PhysRevB.102.155428, PhysRevLett.126.243401, DAlessio2015} that have no direct equilibrium counterparts. Beyond generating novel topological phases, periodic driving provides a versatile platform for controlling transport properties, such as currents and conductance, tuning effective interactions~\cite{PhysRevB.111.125104}, inducing dynamical localization~\cite{PhysRevB.96.144301}, 
generating edge states~\cite{thakurathi2014,seshadri2022}, and engineering synthetic dimensions~\cite{PhysRevX.7.041008, vinjamuri2026mixedfloquetlatticemodel}. These capabilities have opened new avenues for realizing exotic quantum phases, including higher-order topological phases, in 
low-dimensional systems.

Quantum quenches provide another route to study the nonequilibrium dynamics in quantum systems~\cite{annurev:/content/journals/10.1146/annurev-conmatphys-031016-025451, PhysRevB.108.085433}. In a sudden quench, one or more parameters of the system are changed on a timescale much shorter than any intrinsic timescale of the system. The resulting nonequilibrium dynamics can be probed in a variety of experimental platforms~\cite{doi:10.1126/sciadv.1700672, Trotzky2012}. In the context of topological phases, a quantum quench can drive a system from a topologically trivial phase to a topologically nontrivial phase~\cite{PhysRevLett.121.250403}. In such situations, various dynamical observables encode information about the underlying topology of the system~\cite{PhysRevLett.121.250403, PhysRevLett.126.016802, PhysRevB.97.060304, PhysRevResearch.2.033259}. Beyond this, quantum quenches have also been extensively employed to investigate a wide range of nonequilibrium phenomena, including relaxation~\cite{PhysRevB.103.125152}, thermalization~\cite{PhysRevLett.103.056403}, dynamical phase transitions~\cite{PhysRevB.92.104306, PhysRevB.93.144306, PhysRevB.102.060409}, entanglement growth~\cite{PhysRevX.3.031015, PhysRevB.103.085137, PhysRevB.100.125139}, defect generation~\cite{PhysRevB.76.174303}, and the Kibble–Zurek mechanism~\cite{Bacsi2023}.

Several studies have used quantum quenches and periodic driving to investigate the topological phases of quantum systems. The dynamics of edge currents following a quench from a trivial phase to a topological phase has been extensively studied~\cite{PhysRevB.97.115443, PhysRevLett.115.236403}. In the context of open quantum systems, quench dynamics between different nontrivial steady-state phases has been investigated within the framework of Lindblad dynamics~\cite{3wcr-sxtz}. More recently, the long-time behavior of current density after a sudden switch-on of a periodic drive has also been explored~\cite{PhysRevB.93.205437}. Despite these efforts, the nonequilibrium evolution of a quantum system coupled to a bath after the sudden switch-on of a periodic drive remains largely unexplored. In particular, it is still unclear how such a system approaches its steady state and how the signatures of the resulting topological phase manifest in that steady state after the quench.

In this study, we explore the topological signatures of a quantum system following a quantum quench induced by a periodic drive. We consider both a closed system and a system coupled to a bath with a finite coupling strength. Specifically, we study a graphene armchair nanoribbon and compute the bond current and bond conductance near the edges of the nanoribbon, averaged over one driving period in successive time intervals, to characterize the quench dynamics. For the closed system, we find that a finite nonzero dc current, around which the average current oscillates, serves as a signature of the topological phase. When the system is coupled to a bath, the saturation of the average current and conductance indicates the emergence of a steady state, while a finite saturation value of conductance signals the presence of a topological phase. Furthermore, in the limit where the bath behaves as an ideal thermodynamic reservoir, we find that the quantized conductance, obtained after summing over chemical potentials of the bath shifted by integer multiples of the driving frequency, provides information about the number of edge modes in the system.

This paper is organized as follows. In Sec.~\ref{sec:2}, we present the theoretical framework. We begin with a brief overview of Floquet theory in Sec.~\ref{sec:2a}. In Sec.~\ref{sec:2b}, we discuss the density matrix description of the driven system for both the closed-system and system-bath setups. We then outline the calculation of the bond current in both cases and the conductance in the presence of the bath in Sec.~\ref{sec:2c}. Then in Sec.~\ref{sec:2d}, we introduce the model Hamiltonian and its nanoribbon geometry, which serves as the platform for our study.
The results are presented in Sec.~\ref{sec:3}. We first investigate the quench dynamics of the closed system following the sudden switch-on of the drive in Sec.~\ref{sec:3a}. We then examine, in Sec.~\ref{sec:3b},
the emergence of a steady state following the sudden switch-on of a periodic drive when the system is coupled to a bath with a finite coupling strength. Furthermore, we discuss the limit in which the bath acts as an ideal thermodynamic reservoir. Finally, in Sec.~\ref{sec:4}, we summarize our main findings and outline possible directions for future research.

\section{Formalism}
\label{sec:2}
\subsection{Floquet preliminaries}
\label{sec:2a}

For a periodically driven quantum system, the Hamiltonian satisfies the condition \(\mathcal{H}(\bm{k}, t) = \mathcal{H}(\bm{k}, t + T)\),
where \(\bm{k}\) is the Bloch momentum and \(T\) is the driving period. The solutions of the Schr\"odinger equation for such a system satisfy the Floquet theorem
\begin{equation}
\ket{\Psi_{\alpha}(\bm k , t)} = e^{-i\varepsilon_{\bm k, \alpha} t}\ket{\Phi_{\alpha}(\bm k , t)}, \label{eq:A1}
\end{equation}
where \(\varepsilon_{\bm{k},\alpha}\) is the \textit{quasienergy} of the \(\alpha\)-th Floquet band, and \(\ket{\Phi_{\alpha}(\bm{k},t)}\) is the periodic part of the solution that satisfies the condition
\(\ket{\Phi_{\alpha}(\bm{k},t)}=\ket{\Phi_{\alpha}(\bm{k},t+T)}\). (In this paper we will set $\hbar$ to unity unless
stated otherwise).
The quasienergy is uniquely defined within the first \textit{Floquet Brillouin} zone. That is, for each quasienergy \(\varepsilon_{\bm{k},\alpha}\in(-\Omega/2,\Omega/2]\), there exists a unique solution of the form given in Eq.~\eqref{eq:A1}. Following the literature, we call \(\ket{\Psi_{\alpha}(\bm k , t)}\) the \textit{Floquet state} and \(\ket{\Phi_{\alpha}(\bm k , t)}\) the \textit{Floquet mode}, which constitutes a complete orthonormal basis.
Substitution of the Floquet state into the Schr\"odinger equation leads to an eigenvalue equation often referred to as the \textit{Floquet Schr\"odinger equation}
\begin{equation}
\left[\mathcal{H}(\bm k , t) - i\partial_{t}\right]\ket{\Phi_{\alpha}(\bm k , t)} =  \varepsilon_{\bm k, \alpha}\ket{\Phi_{\alpha}(\bm k , t)}.
 \label{eq:A2}
\end{equation}
The evolution of the Floquet state is governed by the evolution operator,
\begin{equation}
\mathcal{U}_{\bm k}(t, t') = \mathcal{T}\exp\left(-i\int_{t'}^{t}\mathcal{H}(\bm k, s)ds\right), \label{eq:A3}
\end{equation}
where \(\mathcal{T}\) is the time-ordering operator. The evolution operator in the Floquet basis can be written as,
\begin{equation}
\mathcal{U}_{\bm k}(t, t') = \sum_{\alpha} e^{-i\varepsilon_{\bm k, \alpha}(t -t')}\ket{\Phi_{\alpha}(\bm k , t)} \bra{\Phi_{\alpha}(\bm k , t')}. \label{eq:A4}
\end{equation}
It follows from the above equation that the evolution operator over one driving period satisfies the eigenvalue equation,
\begin{equation}
\mathcal{U}_{\bm k}(T, 0) \ket{\Phi_{\alpha}(\bm k , 0)} =  e^{-i\varepsilon_{\bm k, \alpha}T}\ket{\Phi_{\alpha}(\bm k , T)}. \label{eq:A5}
\end{equation}
Diagonalizing the above equation gives the quasienergy and the Floquet mode. Owing to the periodicity of the Hamiltonian and the Floquet mode, they can be expressed as a Fourier series
\begin{equation}
\begin{aligned}
\mathcal{H}(\bm k , t) &= \sum_{p \in \mathbb{Z}} e^{-i p \Omega t} \mathcal{H}^{p}(\bm k),\\
\ket{\Phi_{\alpha}(\bm k , t)} &= \sum_{p \in \mathbb{Z}} e^{-i p \Omega t} \ket{\Phi_{\alpha}^{p}(\bm k)}.
\label{eq:A6}  
\end{aligned}
\end{equation}
In the above equation, \(\mathcal{H}^{p}(\bm k)\) and \( \ket{\Phi_{\alpha}^{p}(\bm k)}\) denote the $p$-th Fourier component of the Hamiltonian and the Floquet mode respectively. Substituting Eq.~\eqref{eq:A6} into Eq.~\eqref{eq:A2} leads to the eigenvalue equation (also called \textit{extended zone Hamiltonian}),
\begin{equation}
\sum_{n} \left[\mathcal{H}^{m - n} - m\Omega\delta_{n,m}\right]\ket{\Phi_{\alpha}^{n}(\bm k)} = \varepsilon_{\bm k, \alpha}\ket{\Phi_{\alpha}^{m}(\bm k)}.
\end{equation}
Solving the above eigenvalue equation yields the quasienergy and the Fourier components of the Floquet modes. For a given band $\alpha$, the Fourier components satisfy the normalization condition $\sum_{p} \ket{\Phi_{\alpha}^{p}(\bm k)} \bra{\Phi_{\alpha}^{p}(\bm k)} = \mathbb{I}$.

\begin{figure}[htbp]
%\centering
\includegraphics[width=0.45\textwidth]{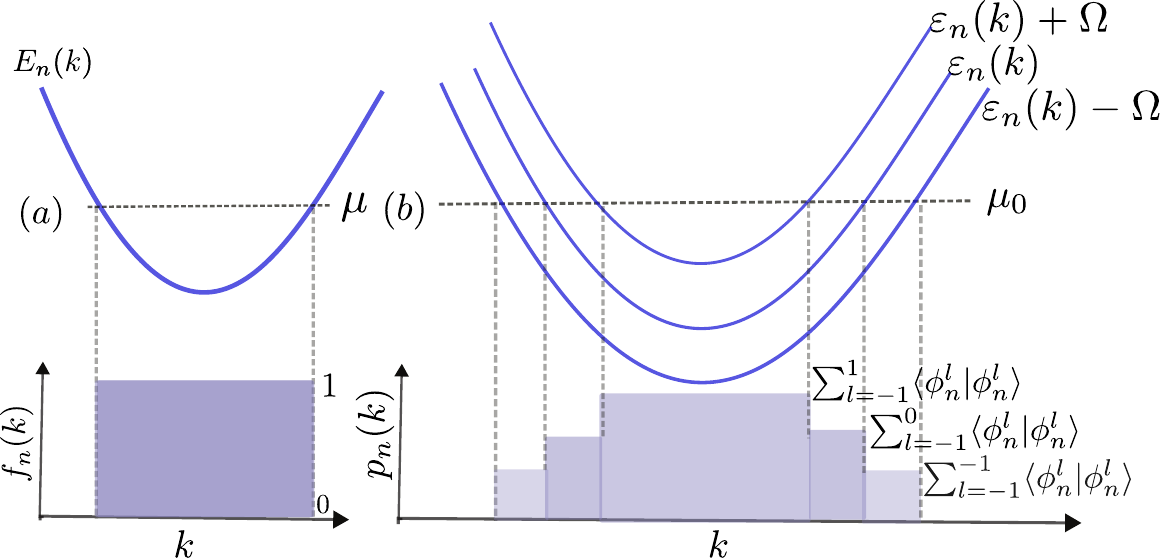}
\caption{(Color online) Figure illustrating the band occupation for both static and periodically driven systems. Panel (a) shows the occupation of the static bands: all states below the chemical potential \(\mu\) are fully occupied (distribution equal to 1), while those above are empty (distribution equal to 0). Panel (b) shows the occupation of the Floquet bands, which exhibits a staircase structure. In contrast to the static case, where the occupation is unity for all filled states, the occupation in the driven system is determined by the weights of the corresponding Floquet copies. As a result, even for states below the chemical potential \(\mu_{0}\), set by the bath, the distribution takes nonzero values, as illustrated in the figure. We consider three Floquet copies to demonstrate the staircase nature of the occupation. Also see Fig. 1 of Ref.~\cite{PhysRevB.107.195135}.}
\label{fig:occupation}
\end{figure}

\subsection{Density matrix of driven quantum system}
\label{sec:2b}

For a periodically driven closed quantum system, in the long-time limit the system either thermalizes to a nontrivial infinite-temperature steady state or retains memory of its initial state if it fails to thermalize. In such systems, several dynamical processes occur. For example, radiative recombination can increase the density of excitations, while photon-assisted electron-phonon scattering can further heat the system~\cite{PhysRevX.5.041050}. The situation changes significantly once the system is coupled to a bath. In the presence of a bath, the system acquires additional channels through which it can remove excess excitations~\cite{KOHLER2005379, PhysRevB.91.184301, PhysRevB.90.195429}. In general, there is a competition between drive-induced dynamical processes and bath-induced relaxation processes. As a result, the system releases into the bath the energy absorbed from the drive, leading to a nontrivial steady state. This steady state is typically reached on a timescale larger than that of the system-bath coupling. Consequently, the density matrix describing the state of the system differs depending on whether the system is closed or coupled to a bath.

In light of the above discussion, we consider two scenarios to describe the density matrix of a periodically driven quantum system. In the first case, the system is closed. In this situation, the density matrix at the onset of the drive can be obtained by projecting the thermal density matrix of the static system onto the Floquet states. The diagonal elements of this density matrix give the projected occupation~\cite{PhysRevB.91.155422, PhysRevB.111.035431} of the Floquet bands, as shown in the equation
\begin{equation}
n^{pr}_{\alpha}(\bm k) =  \bra{\Phi_{\alpha}(\bm k, 0)}\rho^{\rm static}(\bm k)\ket{\Phi_{\alpha}(\bm k, 0)}.
\label{eq:B1}
\end{equation}
Here, $\alpha$ denotes the Floquet band index. The static thermal density matrix is given by
$\rho^{\rm static}(\bm{k}) = \sum_{n} f(E_n(\bm{k})) \ket{n_{\bm{k}}}\bra{n_{\bm{k}}}$,
where $\ket{n_{\bm{k}}}$ is the stationary eigenstate of the static system with energy $E_n(\bm{k})$ corresponding to band $n$. The function $f(x) = 1/(1 + e^{\beta x})$ denotes the Fermi–Dirac distribution function.\\  
In the second case, we consider the driven system to be coupled to a featureless fermionic bath. Such a bath is characterized by a bandwidth much larger than that of the system and by a constant density of states. We further assume that the system can exchange both energy and particles with the bath. Within these approximations, one can solve an effective time-dependent non-Hermitian Schr\"odinger equation, which results in the effective density matrix of the system in the presence of the bath~\cite{PhysRevLett.132.146402} as,
\begin{equation}
\begin{aligned}
\rho_{s}(\bm k, t) &=  \Gamma \int_{-\infty}^{\infty} \frac{d\epsilon}{\pi} f_{0}(\epsilon)\mathcal{U}_{\Gamma}(t,\epsilon)\mathcal{U}_{\Gamma}^{\dagger}(t,\epsilon), \\ 
\mathcal{U}_{\Gamma}(t,\epsilon) &= \int_{-\infty}^{t} dt' \, e^{\Gamma (t' -t)} e^{-i\epsilon t'} \mathcal{U}_{s}(t,t'),
\label{eq:B2}
\end{aligned}
\end{equation}
where $\mathcal{U}_{s}(t,t')$ is the evolution operator of the isolated system, evaluated using Eq.~\eqref{eq:A3} without including the coupling to the bath. The function $f_{0}(\epsilon)$ denotes the Fermi--Dirac distribution function of the bath, which is assumed to be static. Note that Eq.~\eqref{eq:B2} holds for an arbitrary time-dependent system coupled to a static bath.
In the limit of vanishing system-bath coupling, i.e., $\Gamma \to 0$, the featureless fermionic bath effectively behaves as an ideal fermionic bath. In this limit, the time evolution of the density matrix can be evaluated analytically (see Appendix~\ref{app:c} and Ref.~\cite{PhysRevB.107.195135}) and the density matrix
\begin{equation}
\begin{aligned}
\lim_{\Gamma \to 0} \rho_{s}(\bm k, t) &= \sum_{\alpha} p_{\alpha}(\bm k)\ket{\Psi_{\alpha}(\bm k, t)}\bra{\Psi_{\alpha}(\bm k, t)}, \\
p_{\alpha}(\bm k) &= \sum_{l=-\infty}^{\infty} f_{0}(\varepsilon_{\bm k,\alpha}+l\Omega)
\langle \Phi_{\alpha}^{l} | \Phi_{\alpha}^{l} \rangle .
\label{eq:B3}
\end{aligned}
\end{equation}
Here $p_{\alpha}(\bm k)$ is time-independent and represents the \textit{staircase occupation}~\cite{PhysRevB.107.195135, PhysRevLett.133.196601} of the $\alpha$-th Floquet band, as shown in Fig~\ref{fig:occupation}(b).
In the following section, we compute the average edge current and bond conductance using these two different density matrices, corresponding respectively to the cases where the driven system is isolated and where it is coupled to a fermionic bath.

\subsection{Average bond current and bond conductance}
\label{sec:2c}

When a quantum system is subjected to a periodic drive, its physical properties can deviate significantly from those of its static counterpart, irrespective of whether the system is closed or coupled to a bath. For instance, periodic driving can modify the geometry of the bulk band structure, thereby transforming a topologically trivial system into a topologically nontrivial one. Such changes in the topological properties are also reflected in transport measurements such as in the current and conductance.
% At this stage, we compute the current density of the system, which provides information about the chiral edge states when evaluated at the edges of the system.
The current density operator is given by~\cite{PhysRevB.97.115443}
\begin{equation}
\begin{aligned}
\mathcal{J}(t) = \sum_{ij} \mathcal{J}_{ij}(t) = \frac{i}{2}\sum_{ij} \bm r_{ij}\left(t_{ij}e^{\bm r_{ij}\cdot \bm A(t)}\mathcal{C}_{i}^{\dagger}\mathcal{C}_{j} -\rm H.c.\right), \label{eq:C1}
\end{aligned}
\end{equation}
where annihilation operator \(\mathcal{C}_{i}\) destroy the electron at site $i$, \(\bm r_{ij} = \bm r_{i} - \bm r_{j} \), denotes the displacement vector connecting nearest-neighbor sites $i$ and $j$, and \(\bm A(t)\) is the vector potential of the light (here circularly polarized light,  also see the next subsection for details).

For details of the calculation, we refer to Appendix~\ref{app:a}.

At this stage, we compute the current averaged over one driving period within each stroboscopic time interval for the closed system,
\begin{equation}
\begin{aligned}
\Bar{\mathcal{J}}_{A.V}(NT) = \frac{1}{T} \int_{N T}^{(N+1)T} dt \bra{\psi_{g}} \mathcal{U}^{\dagger}(t)\mathcal{J}(t) \mathcal{U}(t) \ket{\psi_{g}}, \label{eq:C2}
\end{aligned}
\end{equation}
where \(\ket{\psi_{g}}\) is the many particle ground state of the static graphene armchair nanoribbon (half filled state), and $N$ is the integer signifying the $N$-th stroboscopic time. For the case of a system coupled to a bath, the same average current in each successive time interval can be computed using
\begin{equation}
\Bar{\mathcal{J}}_{A.V}(NT) = \frac{1}{T} \int_{N T}^{(N+1)T} dt ~\mathrm{Tr} \left[\rho(t)\mathcal{J}(t)\right], \label{eq:C3}
\end{equation}
where \(\rho(t)\) is given by Eq.~\eqref{eq:B2}.
Following the calculation of the average current in each successive time interval, we also compute the conductance of the system when coupled to a bath,
\begin{equation}
\begin{aligned}
\sigma_{A.V}(NT) = \frac{2\pi}{T} \int_{N T}^{(N+1)T} dt ~\mathrm{Tr} \left[\frac{d\rho(t)}{d\mu}\mathcal{J}(t)\right]. \label{eq:C4}
\end{aligned}
\end{equation}
For the calculation details we refer to Appendix~\ref{app:c}. We will study below the bond current and conductance following a sudden 
switch-on of a periodic drive, considering both a closed system and a system coupled to a bath. For numerical calculations, we will compute the current flowing through bonds near or along an edge. 

\begin{figure}[htbp]
%\centering
\includegraphics[width=0.45\textwidth]{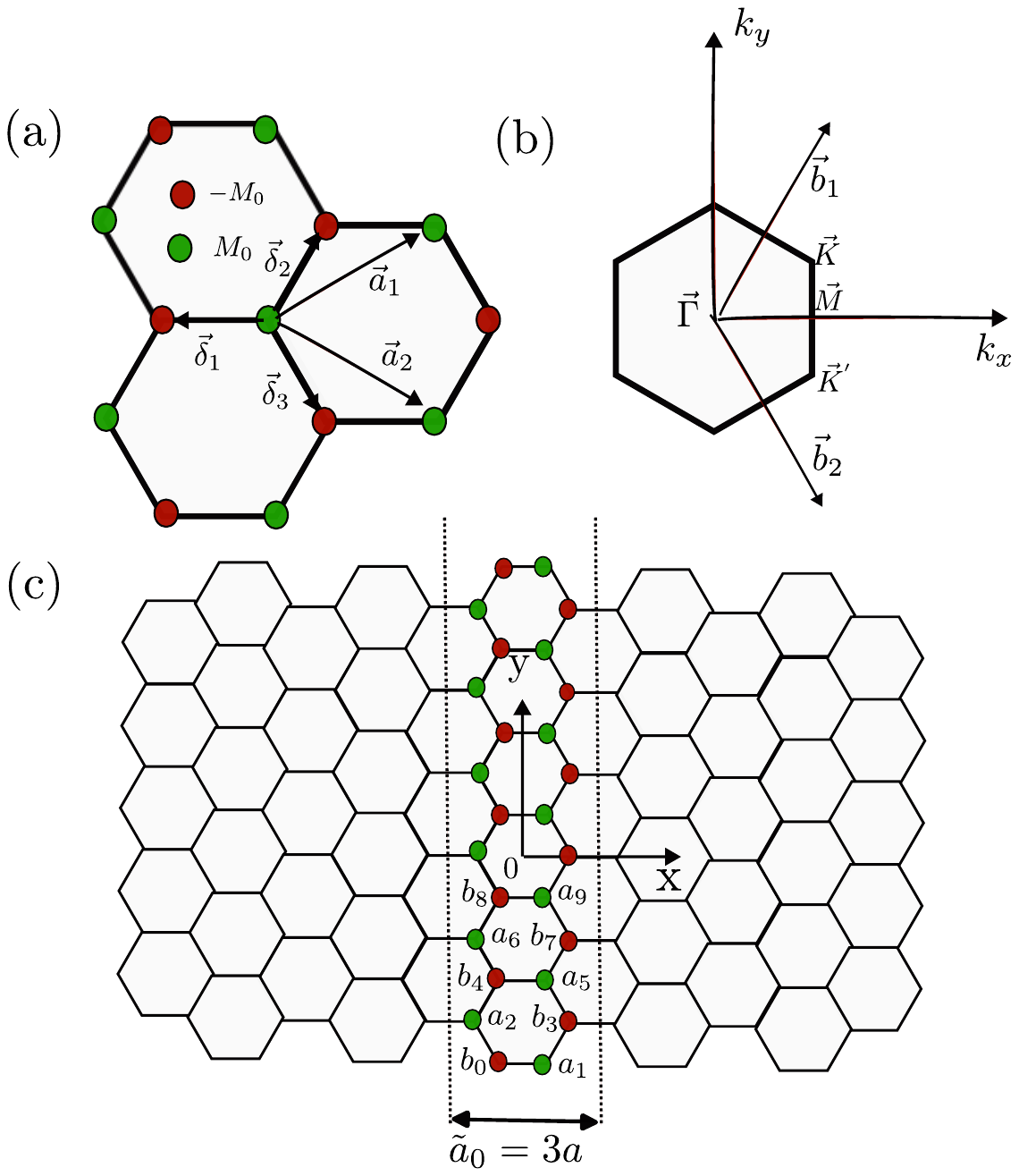}
\vspace{-1.5cm}
\caption{(Color online) (a) Honeycomb lattice of monolayer graphene with a staggered mass $M_0$. The green and red dots represent the sublattices A and B with $M_{0}$. The primitive lattice vectors are $\bm{a}_{1} = a\left(\frac{3}{2}, \frac{\sqrt{3}}{2}\right)$ and $\bm{a}_{2} = a\left(\frac{3}{2}, -\frac{\sqrt{3}}{2}\right)$. 
(b) First Brillouin zone of monolayer graphene. The primitive reciprocal lattice vectors are $\bm{b}_{1} = \frac{2\pi}{3a}(1,\sqrt{3})$ and $\bm{b}_{2} = \frac{2\pi}{3a}(1,-\sqrt{3})$. 
(c) Armchair nanoribbon geometry of the above setup, infinite
along the $x$ direction and finite along the $y$ direction. The lattice constant for the nanoribbon geometry is $\tilde{a}_{0} = 3a$.} 
\label{fig:lattice-geometry}
\end{figure}

\subsection{Model Hamiltonian}
\label{sec:2d}

\begin{figure*}[t]
\centering
\includegraphics[width=1.0\textwidth]{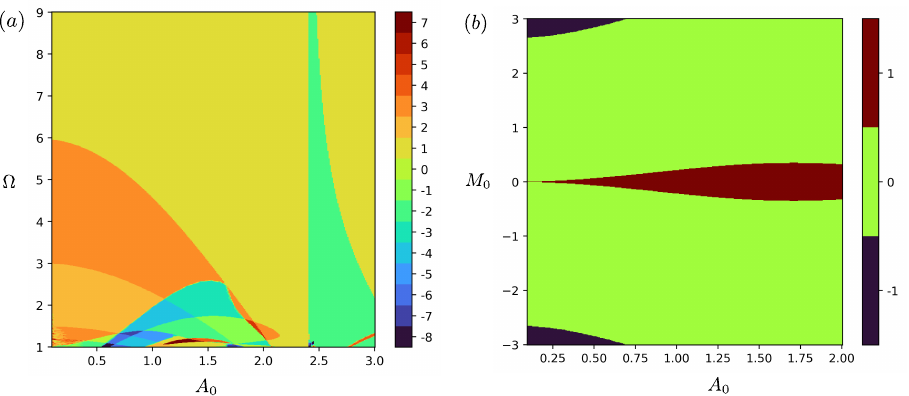}
\caption{(Color online) Chern number phase diagram of periodically driven graphene without (and with) a staggered mass $M_0$. Panel (a) phase diagram as a function of driving frequency and driving amplitude with a staggered mass $M_0 = 0$ . In the low driving frequency regime, the system becomes topologically nontrivial in several regions of the phase space. Higher Chern numbers also appear at low driving frequencies, which arises from the hybridization of different Floquet sidebands with other sidebands or with the original bands. Panel (b) phase diagram as a function of $M_0$ and driving amplitude at $\Omega = 8t_{0}$, where $t_{0}$ is the hopping parameter and we set $t_{0}=1$. The system is topologically nontrivial for small values of the $M_0$ over the entire range of driving amplitudes. For larger values of $M_0$, the system remains topologically nontrivial only for small driving amplitudes.} 
\label{fig:chern_phase}
\end{figure*}

\begin{figure*}[t]
\centering
\includegraphics[width=1.0\textwidth]{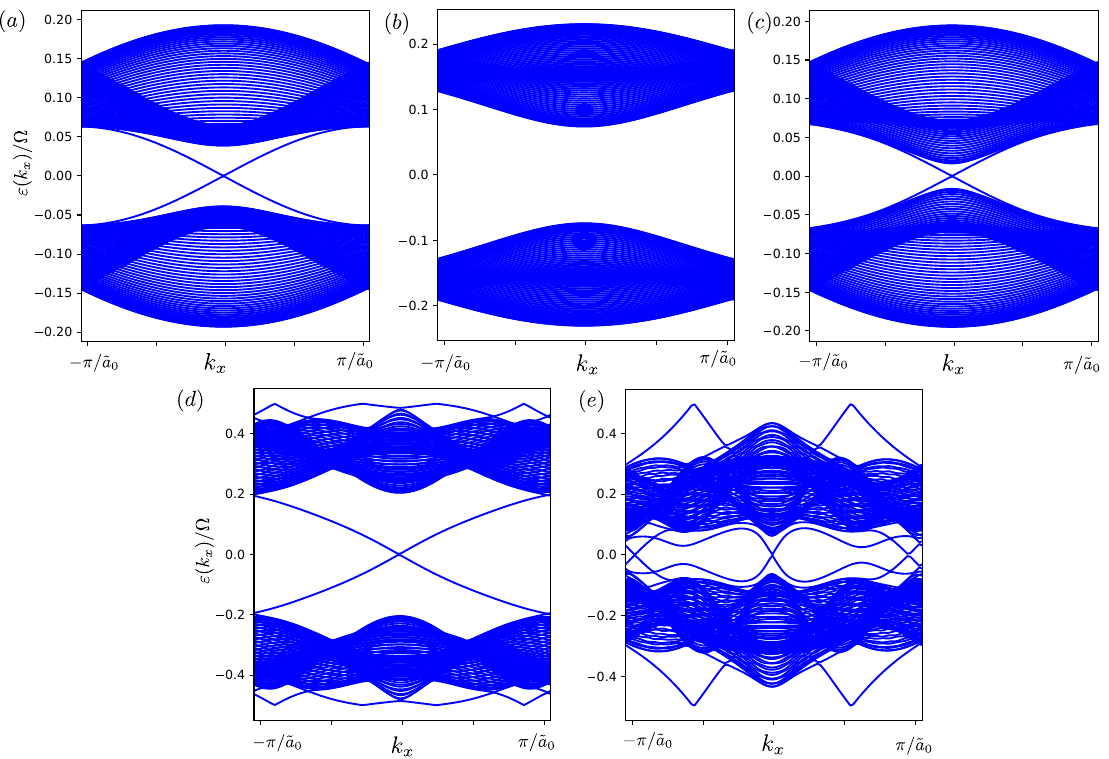}
\vspace{0.3cm}
\caption{(Color online) Quasienergy band structure of periodically driven graphene with (and without) a staggered mass $M_0$. Panels (a)–(c) show the dispersion at high driving frequency, $\Omega = 8t_{0}$, for different values of driving amplitude and $M_0$. The parameters used in these panels are (a) $M_{0} = 0.0,~ A_{0} = 1.5$, (b) $M_{0} = 1.0,~ A_{0} = 1.5$, and (c) $M_{0} = 0.2,~ A_{0} = 1.5$. Panels (d)–(e) show the dispersion at low driving frequencies. The parameters used in these panels are (d)$\Omega = 1.7952t_{0},~M_{0} = 0.0,~ A_{0} = 1.6$ and (e) $\Omega = 1.0472t_{0},~M_{0} = 0.0,~ A_{0} = 1.5$.}  
\label{fig:nanoribbon-disp}
\end{figure*}

To begin with, we consider graphene with a staggered mass $M_0$ periodically driven with circularly polarized light. In the presence of the drive, the time-reversal symmetry of the system breaks, and the $M_{0}$ term breaks the inversion symmetry of the system, making the system topologically nontrivial. The time-dependent Bloch Hamiltonian of the system is given by
\begin{equation}
\mathcal{H}(\bm k , t) = 
\begin{pmatrix}
M_{0} && g( \bm k , t) \\
g( \bm k , t)^{*} && -M_{0}
\end{pmatrix},
\end{equation}
where \(g( \bm k , t)= t_{0}e^{i( \bm k \cdot \bm \delta_{1} + \bm A(t) \cdot \bm \delta_{1})} + t_{0} e^{i( \bm k \cdot \bm \delta_{2} + \bm A(t) \cdot \bm \delta_{2})} + t_{0} e^{i( \bm k \cdot \bm \delta_{3} + \bm A(t) \cdot \bm \delta_{3})} \), with $t_{0}$ being the hopping amplitude of the nearest neighbor. Here, \(\bm \delta_{1} = a(-1, 0), \bm \delta_{2} = a(\frac{1}{2}, \frac{\sqrt{3}}{2}) \), \(\bm \delta_{3} = a(\frac{1}{2}, -\frac{\sqrt{3}}{2})\) are three nearest-neighbor vectors and $a$ is the lattice constant. The effect of the drive (circularly polarized light) is incorporated into the Hamiltonian through the Peierls substitution, where the momentum $\bm{k}$ is replaced by $\bm{k} + \bm{A}(t)$. Here $\bm{A}(t) = A_{0}(\cos{\Omega t}, \sin{\Omega t})$ is the vector potential of circularly polarized light, with $A_{0}$ being the driving amplitude and $\Omega$ the driving frequency. The competition between the effects of the drive and the staggered mass leads to a rich topologically nontrivial phase diagram. Using Floquet theory as discussed above, one can compute the quasienergy band structure and the corresponding Berry curvature. The integration of the Berry curvature of the quasienergy bands over the entire Brillouin zone gives the Chern number, as discussed in Sec.~\ref{sec:3}. 

In order to compute the bond current and bond conductance, we consider the armchair nanoribbon geometry of the above setup. We choose armchair edges because they naturally open a band gap in the undriven system, giving a well-defined trivial ground state before the quench; zigzag nanoribbons, in contrast, host gapless edge states even without driving. The corresponding lattice geometry is shown in Fig.~\ref{fig:lattice-geometry}(c). We consider the lattice to be infinite along the $x$ direction and finite along the $y$ direction. The time-dependent Bloch Hamiltonian for the driven armchair graphene nanoribbon is given by
\begin{equation} 
\begin{aligned}
\Tilde{\mathcal{H}}( k_{x} , t) &= t_0 \sum_{j = 0}^{N_{\mathrm{hex}}} \exp(i\bm \delta_{1} \cdot \bm A(t)) a_{4j+1}^{\dagger}b_{4j}\\
&+ t_0 \sum_{j = 0}^{N_{\mathrm{hex}}-1} [ \exp(i\bm \delta_{1} \cdot \bm A(t)) \exp(i\tilde{a}_{0} k_{x}) a_{4j+2}^{\dagger}b_{4j+3}\\
&+ t_{0} \exp(i\bm \delta_{2} \cdot \bm A(t)) (a_{4j +1}^{\dagger}b_{4j+3}+ a_{4j +2}^{\dagger}b_{4j+4})\\
&+ t_{0} \exp(i\bm \delta_{3} \cdot \bm A(t)) (a_{4j +2}^{\dagger}b_{4j} + a_{4j +5}^{\dagger}b_{4j + 3})] \\
&+ \rm H.c.,
\end{aligned}
\end{equation}
where \(k_{x}\) is the good quantum number (crystal momentum) in the $x$ direction. $N_{\mathrm{hex}}$ is the number of hexagons in the finite direction of the nanoribbon, and $a_k$ ($b_k$) denotes the annihilation operator for an electron in the $A$ ($B$) sublattice, where $k$ denotes the lattice index, and \(\tilde{a}_{0}\) is the lattice constant of the geometry of the armchair nanoribbon (also see Fig.~\ref{fig:lattice-geometry}). In order to take into account the effect of a staggered mass we add the term \( M_0 \sum_{j} (a_{j}^{\dagger}a_{j} - b_{j}^{\dagger}b_{j}) \) to the above Hamiltonian. The quasienergy dispersion of the driven armchair graphene nanoribbon is shown for various values of the driving amplitude and $M_0$ in Fig.~\ref{fig:nanoribbon-disp}. 

\section{Results and Discussion}
\label{sec:3}

When a quantum system is subjected to a periodic drive, such as circularly polarized light, time-reversal symmetry is broken, leading to gap openings and a renormalization of the band structure. As a result, the system can host nontrivial topological phases. In this work, we investigate the signatures of these topological phases following a quench with a periodic drive, considering both isolated systems and systems coupled to a bath.

We consider graphene irradiated by circularly polarized light and compute the phase diagram of the Chern number of the lower Floquet band as a function of the driving amplitude and frequency, as shown in Fig.~\ref{fig:chern_phase}(a). In periodically driven systems, the Chern number of a Floquet quasienergy band is determined by the difference between the number of chiral edge modes entering and leaving the band. The number of edge modes crossing a given quasienergy gap can be obtained from the winding of the micromotion operator around that quasienergy~\cite{PhysRevX.3.031005, PhysRevLett.113.236803}. In the high-frequency regime, the periodic drive opens gaps at the gapless points of the undriven graphene spectrum, with the gap size depending on the driving strength. In this regime, different Floquet sidebands remain well separated and do not overlap. Consequently, chiral edge modes appear only across the zero quasienergy gap. In contrast, in the low-frequency regime, Floquet sidebands hybridize due to band folding at the Floquet zone boundary. As a result, additional edge modes emerge both across the zero-quasienergy gap and at the Floquet zone boundary. The edge modes appearing at the Floquet zone boundary are commonly referred to as anomalous edge states~\cite{PhysRevX.3.031005}.

To explicitly demonstrate the presence of edge modes, we plot the quasienergy dispersion of driven graphene in the armchair nanoribbon geometry, as shown in Fig.~\ref{fig:nanoribbon-disp}. The appearance of edge states, including anomalous edge states, shows the existence of nontrivial topological phases. Importantly, in periodically driven systems it is possible to have a zero Chern number while still supporting edge modes. This occurs when the number of edge modes entering and leaving a band are equal. Such phases, which have no equilibrium counterpart, are characterized by a nonzero winding number of the micromotion operator~\cite{PhysRevX.3.031005}. We further investigate the effect of an on-site mass term in driven graphene. The competition between the effective mass generated by the periodic drive and the on-site mass term significantly modifies the phases of the system. In Fig.~\ref{fig:chern_phase}(b), we show the Chern number phase diagram as a function of on-site mass term and driving amplitude for a fixed driving frequency $\Omega = 8t_{0}$. Depending on the driving parameters, namely the driving amplitude and driving frequency, the on-site mass term can either destroy or preserve the edge states. In Fig.~\ref{fig:nanoribbon-disp}(b) and Fig.~\ref{fig:nanoribbon-disp}(c), we plot the nanoribbon quasienergy dispersion in the armchair geometry to demonstrate, respectively, the disappearance and survival of edge modes in the presence of the on-site mass term.

In this work, we explore the signatures of topological phases in periodically driven graphene by analyzing the quench dynamics of an armchair nanoribbon, considering both closed systems and systems coupled to an external bath.

\subsection{Quench dynamics in periodically driven closed graphene nanoribbons}
\label{sec:3a}

\begin{figure*}[t]
\centering
\includegraphics[width=1.05\textwidth]{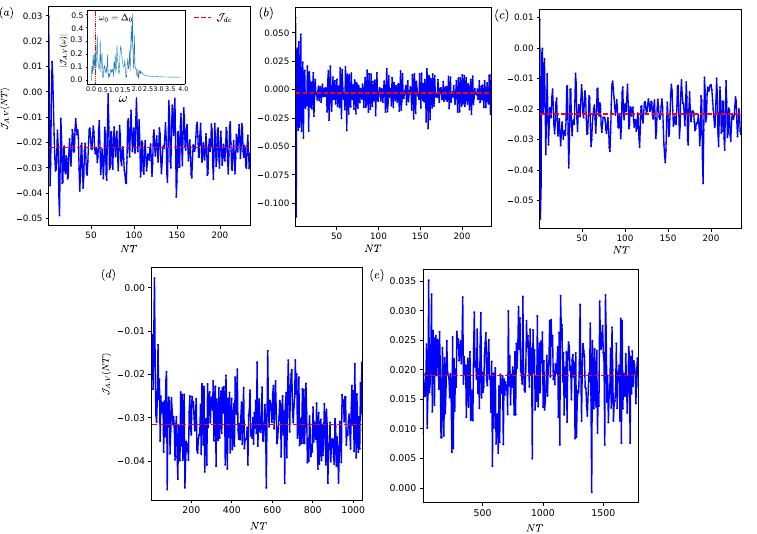}
\vspace{0.2cm}
\caption{(Color online) Average bond current near the edge over one driving period in each successive time interval after the onset of the periodic drive. The current exhibits strong temporal oscillations around a finite dc value (red dashed line). This dc value is given by the weighted sum of the average current over one driving period in a Floquet band, with weights determined by the projected occupations of that Floquet band. A nonzero dc current indicates a topological phase of the driven graphene nanoribbon. The upper panels correspond to high driving frequency, $\Omega = 8t_{0}$. Panel (a) $M_{0} = 0.0,~ A_{0} = 1.5$ and (c) $M_{0} = 0.2,~ A_{0} = 1.5$ correspond to the topological phase, while panel (b) $M_{0} = 1.0,~ A_{0} = 1.5$ corresponds to the trivial phase. The inset in panel (a) shows the frequency-resolved current $|\mathcal{J}_{A.V}(\omega)|$, which exhibits a kink at $\omega=\Delta_{0}$, where $\Delta_{0}=0.0369\Omega$.  Panels (d) and (e) show the current in the topological phase at lower driving frequency. The parameters used in these panels are: panel (d) $\Omega = 1.7952t_{0},~M_{0} = 0.0,~ A_{0} = 1.6$ and (e) $\Omega = 1.0472t_{0},~M_{0} = 0.0,~ A_{0} = 1.5$. Further the magnitude of the dc value in the topological phase depends upon the gap induced by the drive and the driving parameters. In all panels we take $N_{b} = 1$.}
\label{fig:proj-curr}
\end{figure*}

In this section, we study the dynamics of a closed armchair graphene nanoribbon following the sudden onset of circularly polarized light. To probe the evolution of the system, we calculate the current averaged over one driving period using contributions from a few bonds near the edge at successive stroboscopic times.

The current operator $\mathcal{J}(t)$ near the edge is given by
\begin{equation}
\mathcal{J}(t) = \sum_{i = 0}^{i = N_{b}} \mathcal{J}_{4i , 4i+1}(t),
\label{eq:ra1}
\end{equation}
where $\mathcal{J}_{ij}(t)$ is the bond current given in Eq.~\eqref{eq:C1}. $N_{b}$ is the number of bonds parallel to the nanoribbon edge in the transverse direction (perpendicular to the edge).
%Note in the case, when either of the edge state is dispersed deep in the bulk
Upon a sudden switch-on of the periodic drive, we find that the current oscillates around a constant dc value, $\mathcal{J}_{\mathrm{dc}}$, as shown by the red dotted line in Fig.~\ref{fig:proj-curr}. 
The dc current can be extracted from the expression for the average current over one driving period given in Eq.~\eqref{eq:C2} by rewriting it as
\begin{equation}
\begin{aligned}
\Bar{\mathcal{J}}_{A.V}(NT) &= \frac{1}{ T}\int_{0}^{ T} \bra{\psi_{g}}\mathcal{U}^{\dagger}(t' + NT)\mathcal{J}(t' + NT)\\
& ~~~~~~~~~~~~~\times\mathcal{U}(t' + NT )\ket{\psi_{g}} dt'.
\end{aligned}
\end{equation}
Owing to the noninteracting nature of the static system, we express the expectation value with respect to many particle ground state of the static Hamiltonian (half filled state) in terms of sum of expectation value with respect to all occupied single particle eigenstate,
\begin{equation}
\begin{aligned}
\Bar{\mathcal{J}}_{A.V}(NT) &= \sum_{n \in occ}\frac{1}{ T}\int_{0}^{ T} \bra{n}\mathcal{U}^{\dagger}(t' + NT)\mathcal{J}(t' + NT)\\
& ~~~~~~~~~~~~~\times\mathcal{U}(t' + NT )\ket{n} dt'.
\end{aligned}
\end{equation}
In the above equation, ``occ" stands for the occupied state, \(\ket{n}\) denotes the single particle eigenstate of the static Hamiltonian.
At zero temperature, using the Fermi function, the sum over occupied single particle eigenstate can be cast as sum over all eigenstate,
\begin{equation}
\begin{aligned}
\Bar{\mathcal{J}}_{A.V}(NT) &= \sum_{n}\frac{1}{ T}\int_{0}^{ T} f(E_{n})\bra{n}\mathcal{U}^{\dagger}(t' + NT)\mathcal{J}(t' + NT)\\
& ~~~~~~~~~~~~~\times\mathcal{U}(t' + NT )\ket{n} dt',
\end{aligned}
\end{equation}
where $f(x)$ denotes the Fermi function and $E_{n}$ denotes the corresponding eigenvalue for the $n$-th band of the static system. The details of the calculation are provided in Appendix~\ref{app:b}. 
 Further, using the periodicity of the current operator and by expressing the evolution operator in the Floquet basis, the expression for the average current can be written, after some algebra, as :
\begin{equation}
\begin{aligned}
\Bar{\mathcal{J}}_{A.V} &= \frac{1}{ T}\int_{0}^{T} \mathrm{Tr}\sum_{n, \alpha, \beta} e^{i N t'(\varepsilon_{\alpha} - \varepsilon_{\beta})} f(E_{n})\ket{n}\langle n\ket{\Phi_{\alpha}(0)}\\
& ~~~~~~~~~~~~\times \bra{\Phi_{\alpha}(N t')}
\mathcal{J}(N t')\ket{\Phi_{\beta}(N t')}\bra{\Phi_{\beta}(0)} dt',
\end{aligned}
\end{equation}
 For large \(N\), we retain only the \(\alpha=\beta\) terms, since the \(\alpha\neq\beta\) terms oscillate with different frequencies for different values of the momentum \(k_x\). Consequently, the summation over \(k_x\) leads to the vanishing of the \(\alpha\neq\beta\) terms due to dephasing. As a result, the above expression simplifies to a constant dc value,
\begin{equation}
\begin{aligned}
\mathcal{J}_{dc} &= \sum_{k_{x}}\frac{1}{ T}\int_{0}^{T}\sum_{n, \alpha}  f(E_{n})\\
 & ~~~~~~\times |\braket{n |\Phi_{\alpha}(0)}|^{2}
\bra{\Phi_{\alpha}(N t')}
\mathcal{J}(N t')\ket{\Phi_{\alpha}(N t')} dt'.
\end{aligned}
\end{equation}
 After rearranging the above expression, the dc current is given by the sum of the average current over one driving period associated with each Floquet band, weighted by the occupation of the corresponding Floquet band,
\begin{eqnarray}
\mathcal{J}_{dc} &=& \sum_{\alpha, k_{x}} n_{\alpha}^{pr}(k_{x}) \Bar{\mathcal{J}}_{\alpha}(k_{x}), \nonumber \\
\Bar{\mathcal{J}}_{\alpha}(k_{x}) &=& \frac{1}{T}\int_{0}^{T}  \bra{\Phi_{\alpha}(k_{x} , t)}\mathcal{J}_{\mathrm{tot}}(t) \ket{\Phi_{\alpha}(k_{x} , t)} dt,
\end{eqnarray} 
where $n_{\alpha}^{pr}(k_{x})$ denotes the projected occupation of the $\alpha$-th Floquet band at momentum \(k_{x}\) given by Eq.~\eqref{eq:B1}. The projected occupation dictates the electron distribution in the quasienergy band just after the sudden quench of the periodic drive~\cite{PhysRevB.111.035431}. A nonzero value of $\mathcal{J}_{\mathrm{dc}}$ signifies the topological phase of the driven system and indicates the presence of edge modes in the graphene nanoribbon as shown in Figs.~\ref{fig:proj-curr}(a,c,d,e), while a vanishingly small value of the dc current corresponds to the trivial phase as shown in Fig.~\ref{fig:proj-curr}(b). The magnitude of $\mathcal{J}_{\mathrm{dc}}$ depends on the driving protocol as well as the size of the bulk gap. The distinction between the topological and trivial phases is more pronounced through the $\mathcal{J}_{dc}$ current, if edge states localize more near the edge of the nanoribbon. We find that, although the current flows in opposite directions along the two edges of the nanoribbon in the topological phase, a finite net dc current remains. This is consistent with~\cite{PhysRevB.93.205437}, where it is shown that a laser switch-on protocol following a quench breaks the inversion symmetry of occupation distribution. Together with the broken time-reversal symmetry of the Floquet Hamiltonian, this results in asymmetric quasienergy-band occupations and hence a nonzero edge-current density in the long-time limit. In this study, we compute the current averaged over one driving period for long evolution times encompassing a large number of stroboscopic cycles. We find that the current continues to oscillate around the same dc value observed at short times, with no noticeable drift even in the long-time limit. This behavior reflects the  indefinite persistence of a non-thermal state in a closed, noninteracting integrable periodically driven system~\cite{10.21468/SciPostPhys.2.3.021, PhysRevB.100.100302}. In the absence of interactions, the system cannot thermalize and thus retains memory of the initial state. In real materials, residual interactions or coupling to phonons would eventually destroy this non-thermal state, but the topological signatures described here should be observable on experimentally accessible timescales before heating becomes significant. Consequently, the system retains memory of its initial state and associated current, making the topological signatures of the quench dynamics accessible to experimental observation. To gain insight into the current oscillations following the quench, we perform a discrete Fourier transform of the time-dependent current. We find that the observed peaks correspond to transitions between different quasienergy levels. In the inset of Fig.~\ref{fig:proj-curr}(a), we explicitly highlight the peak associated with the bulk quasienergy gap~$\Delta_{0}$. We also compute the average current for a very low driving frequency, at which anomalous edge modes traversing the Floquet zone boundary along with edge modes traversing the zero quasienergy are present, as shown in Fig.~\ref{fig:proj-curr}(e).

Further, our study highlights a fundamental difference between quenches involving only static Hamiltonians and quenches from a static to a periodically driven regime. In the former case, the current oscillates around a dc value determined by the equilibrium properties of the final Hamiltonian, specifically its ground state~\cite{PhysRevB.97.115443,PhysRevLett.115.236403}. In contrast, following a quench from a static Hamiltonian to a periodically driven Hamiltonian, the dc current is governed by the occupations of the Floquet states and the current associated with those states. These occupations are determined by the overlap between the ground state of the initial Hamiltonian and the Floquet mode at the moment the drive is switched on.

\begin{figure*}[t]
\centering
\includegraphics[width=0.8\textwidth]{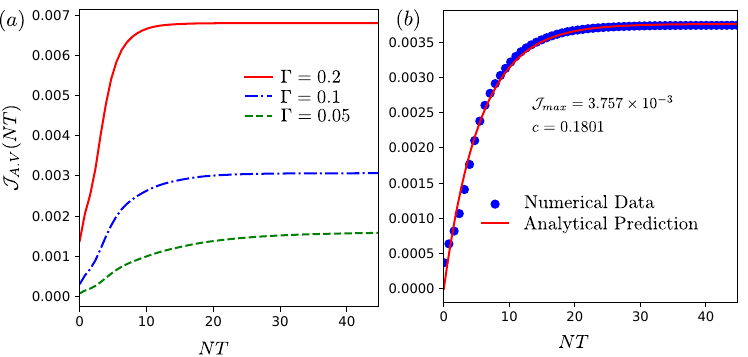}
\vspace{0.2cm}
\caption{(Color online) Average current over one driving period in each successive time interval near the edge after the onset of the periodic drive when system is coupled to a bath via a single coupling scale $\Gamma$.  Panel (a) corresponds to the average current for various values of coupling scale. The other parameters used in this plot are  $\Omega = 8t_{0},~A_{0}=1.5$, $~M_{0}=0$. Panel (b) shows the agreement between average current obtained via numerics and the predicted analytical behavior of current as a function of time. The parameters used in this plot are $\Omega = 8t_{0},~A_{0}= 1.5$, $~M_{0}=0,~\Gamma=0.1$. 
}
\label{fig:ss-curr}
\end{figure*}

\subsection{Quench dynamics in driven graphene nanoribbon coupled to a bath}
\label{sec:3b}

\begin{figure*}[t]
\centering
\includegraphics[width=0.8\textwidth]{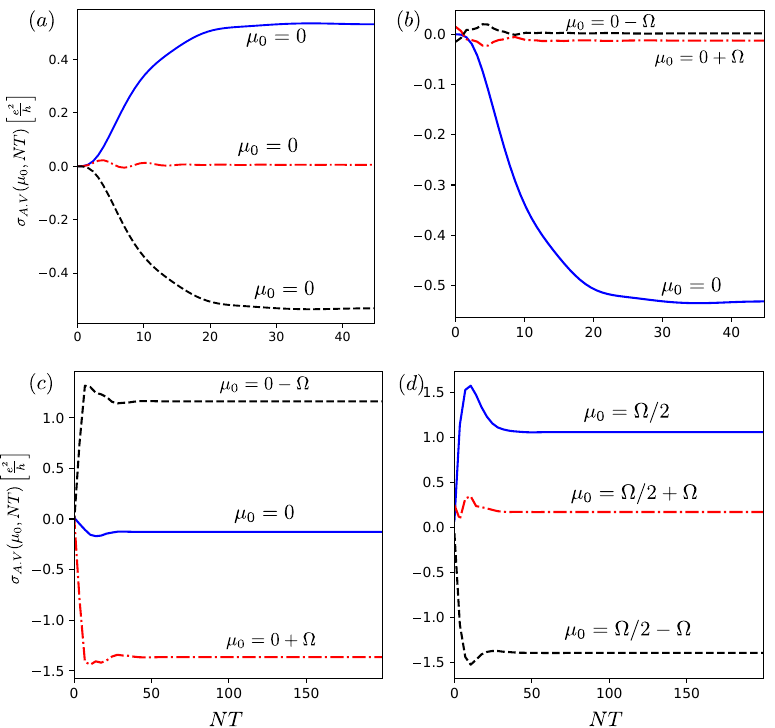}
\vspace{0.2cm}
\caption{(Color online) Average bond conductance near the edge over one driving period in each successive time interval after the onset of the periodic drive when the system is coupled to a bath via a single coupling scale $\Gamma$. The upper panels correspond to a high driving frequency, $\Omega = 8t_{0}$. Panel (a) demonstrates the bond conductance near the two edges in the topological phase (using blue solid and black dashed lines) and in the trivial phase (using red dashed-dotted line). Panel (b) demonstrates the average conductance with bath chemical potential \(\mu_{0} = 0, \pm\Omega\). The other parameters are $M_{0} = 0.0,~ A_{0} = 1.5$. Panel (c) demonstrates the average conductance at low driving frequencies with bath chemical potential \(\mu_{0} = 0, \pm\Omega\). The parameters used in this plot are $\Omega = 1.7952t_{0},~M_{0} = 0.0,~ A_{0} = 1.6$. Panel (d) demonstrates the average conductance for same parameter values given in panel (c) except with the bath chemical potential placed at \(\mu_{0} = \Omega/2, \Omega/2 \pm\Omega\).}
\label{fig:cond}
\end{figure*}

The behavior of the system changes significantly upon coupling to an external bath. We consider a graphene nanoribbon coupled to a \textit{featureless fermionic bath}, characterized by a single effective coupling strength $\Gamma$. For a static system, finite coupling to the bath leads to deviations from the equilibrium Fermi-Dirac distribution. As $\Gamma \to 0$, the bath behaves as an \textit{ideal thermodynamic reservoir} and the electronic occupation recovers the Fermi-Dirac distribution form. In contrast, a periodically driven system in the presence of a bath always deviates from following a Fermi-Dirac distribution. Even in the limit $\Gamma  \to 0$, a driven system in its steady state phase exhibits a staircase structure, as studied in Ref.~\cite{PhysRevB.107.195135} (also see Fig.~\ref{fig:occupation}).

Owing to the noninteracting nature of the combined system and bath, the Schr\"odinger equation can be solved  exactly, the only simplification being the assumption of a single relaxation scale, $\Gamma$~\cite{PhysRevB.107.195135}. By integrating out the bath degrees of freedom, one obtains an effective density matrix of the system given by Eq.~\eqref{eq:B2}.

In this section, we study the quench dynamics of a periodically driven graphene nanoribbon coupled to a bath. As in the case of a closed driven system, we characterize the dynamics through the average bond current near the edge. The current operator is given by Eq.~\eqref{eq:ra1}. Following the quench, the current increases with time before reaching a saturation value. The corresponding timescale is  inversely proportional to the system-bath coupling strength \(\Gamma\), as illustrated in Fig.~\ref{fig:ss-curr}(a).
The time evolution of the current is well captured by
\begin{equation}
\mathcal{J}(t)=\mathcal{J}_{\mathrm{max}}\left(1-e^{-ct}\right),
\end{equation}
where \(c\) is a fitting parameter and \(\mathcal{J}_{\mathrm{max}}\) is the saturated current. The extracted values of \(\mathcal{J}_{\mathrm{max}}\) are shown as a solid red line in Fig.~\ref{fig:ss-curr}(b), which is in good agreement with our numerical result shown as a dotted line in the same figure.
The saturation of the average current indicates that the system reaches a unique steady state for a given system-bath coupling strength.

In the presence of a bath, the conductance is a more relevant quantity than the current, as it provides direct information about the nature of the edge modes and the bulk topological invariant of the system. For example, in a multiterminal setup, a topological system exhibits a quantized tunneling conductance, which serves as a signature of the underlying edge states~\cite{PhysRevLett.113.266801}.
In this study, analogous to the average current discussed above, we compute the average bond conductance near the edge of the nanoribbon over one driving period as a function of the stroboscopic time using Eq.~\eqref{eq:C4}. After some calculation it further reduces to
\begin{eqnarray}
&& \sigma_{A.V}(\mu_{0} , NT) \nonumber \\
&& = \frac{2\pi}{T} \int_{N T}^{(N+1)T} \mathrm{Tr} \left[\mathcal{U}_{\Gamma}(t, \mu_{0})\mathcal{J}(t)\mathcal{U}^{\dagger}_{\Gamma}(t, \mu_{0})\right] dt,
\end{eqnarray}
where $\mathcal{U}_{\Gamma}(t,\mu_{0})$ is given by Eq.~\eqref{eq:B2}. For details of the calculation we refer to Appendix~\ref{app:c}.
We find that the average conductance initially increases with time and subsequently saturates after a characteristic timescale. Moreover, the average conductance at opposite edges has opposite signs, revealing the chiral nature of the edge mode in the topological phase, as shown in Fig.~\ref{fig:cond}(a). The saturated value of the average conductance is significantly larger in the topological phase than in the trivial phase. This behavior indicates that the topological phase of the system enhances transport in the presence of a bath.

We further emphasize that, in the high-frequency driving regime, a finite conductance is obtained in the topological phase when the bath chemical potential is fixed at zero. However, the conductance vanishes when the bath chemical potential is shifted by an integer multiple of the driving frequency. To demonstrate this behavior, we compare the average conductance evaluated at zero bath chemical potential with that obtained when the chemical potential is shifted by integer multiples of the driving frequency, as shown in Fig.~\ref{fig:cond}(b). In contrast, in the low-frequency driving regime, the average conductance corresponding to bath chemical potentials shifted by integer multiples of the driving frequency remains finite and saturates to different values. This behavior is observed both when the bath chemical potential is placed at zero quasienergy and when it is located at the Floquet zone boundary, as illustrated in Fig.~\ref{fig:cond}(c,d). The difference originates from the increasingly pronounced staircase structure of the nonequilibrium distribution function at lower driving frequencies. To gain further insight into the physical significance of the conductance evaluated at different bath chemical potentials, we now discuss the limit \(\Gamma \to 0\).

In the limit $\Gamma \to 0$, the system requires an infinitely long time, of $\mathcal{O}(1/\Gamma)$, to reach the steady state. In this regime, the occupations of the Floquet quasienergy bands are described by Eq.~\eqref{eq:B3}, and the bath effectively behaves as an ideal thermodynamic bath. We calculate the total average bond conductance near both edges of the nanoribbon by summing the average bond conductance over bath chemical potentials shifted by integer multiples of the driving frequency. Specifically, we consider bath chemical potentials located near zero quasienergy and near the Floquet zone boundary. The resulting total average conductance, at zero temperature, at the edges of the nanoribbon, expressed in units of $e^{2}/h$ (see Appendix~\ref{app:c} for the calculation details), is given by
\begin{equation}
\begin{aligned}
\sigma_{ss}^{i}(\mu_{0}) &= \frac{1}{2\pi}\int d\bm k \sum_{\alpha} \sum_{\ell,m=-\infty}^{\infty}
\delta\!\left(\varepsilon_{\bm k,\alpha} + \ell\Omega - \mu_{0} - m\Omega\right)\\
&\times \langle \Phi_{\alpha}^{l} | \Phi_{\alpha}^{l} \rangle\frac{1}{T} \int_{0}^{T}
\bra{\Phi_{\alpha}(\bm k, t)} J(t) \ket{\Phi_{\alpha}(\bm k, t)} dt,
\end{aligned}
\end{equation}
where $i = {l,u}$ denotes the lower and upper edge respectively. We find that the quantity $\tilde{\sigma}(\mu_{0}) = (\sigma_{ss}^{u}(\mu_{0}) - \sigma_{ss}^{l}(\mu_{0}))/2 $ approaches quantization and hence gives the information of the number of edge mode crossing the chemical potential $\mu_{0}$. Further, the signs of $\sigma_{ss}^{u}(\mu_{0})$ and $\sigma_{ss}^{l}(\mu_{0})$ give the information of the chirality of the edge modes at the upper or lower edges of the nanoribbon crossing the chemical potential $\mu_{0}$. In Table~\ref{tab:summed_conductance}, we tabulate the value of the total average conductance for various values of the driving frequency and driving amplitude, around the chemical potential $\mu_{0} = 0$ and $\mu_{0} = \Omega/2$. Our results are consistent with a recent study~\cite{kumari2026quantizedtransportfloquettopological} 
which employed the Floquet nonequilibrium Green's function (NEGF) formalism to demonstrate, using exact numerical calculations for a strip geometry, that the two-terminal (longitudinal) conductance is quantized as $|W_{\epsilon}|e^{2}/h$. Here, $W_{\epsilon}$ denotes the winding number of the micromotion operator at quasienergy $\epsilon$, which is equal to the number of edge modes crossing the quasienergy gap at $\epsilon$. To further support our findings, we compute the two-terminal conductance of a finite-size tight-binding model of graphene driven by a three-step driving protocol (defined in Appendix~ \ref{app:d}). This protocol closely mimics circularly polarized light by breaking time-reversal symmetry while simultaneously opening a large bulk gap, thereby enabling reliable transport calculations in a two-terminal geometry. We further calculate the corresponding bond conductance and find that, after applying the sum rule, it is consistent with our results. Additional details are provided in Appendix~\ref{app:d}.

\renewcommand{\arraystretch}{1.9}
\setlength{\tabcolsep}{10pt}
\begin{table*}[t]
\centering
\begin{tabular}{|c|c|c|c|c|c|c|c|}
\hline
$A_{0}$ & $\Omega$ & $\sigma_{ss}^{u}(\mu_{0} = 0)$ & $\sigma_{ss}^{l}(\mu_{0} = 0)$ & $\tilde{\sigma}_{ss}(\mu_{0} = 0)$ & $\sigma_{ss}^{u}(\mu_{0} = \Omega/2)$ & $\sigma_{ss}^{l}(\mu_{0} = \Omega/2)$ & $\tilde{\sigma}_{ss}(\mu_{0} = \Omega/2)$ \\
\hline
$1.6$ & $1.7952$ & $-0.9993$ & $1.0004$ &
$0.9999$ & $-3.9872$ & $3.9961$ & $3.9916$ \\
\hline
$1.3$ & $1.3963$ & $0.0$ & $0.0$ &
$0.0$ & $-2.0009$ & $1.9984$ & $1.9997$ \\
\hline
$1.5$ & $1.0472$ & $-2.9002$ & $2.9226$ &
$2.9114$ & $-2.0003$ & $1.9994$ & $1.9998$ \\
\hline
$1.5$ & $8$ & $-0.9998$ & $1.0001$ &
$0.9999$ & $0.0$ & $0.0$ & $0.0$ \\
\hline
$1.5$ & $5$ & $-0.9998$ & $1.0001$ &
$0.9999$ & $0.0$ & $0.0$ & $0.0$ \\
\hline
\end{tabular}
\vspace{0.5cm}
\caption{Table showing the total average conductance near both (upper and lower) edges of nanoribbon with the chemical potential of the bath \(\mu_{0}\) placed at zero quasienergy and Floquet zone boundary. For the purpose of numerical convergence, we consider all the bonds from edge to middle of the nanoribbon. We use the symbol $\sigma_{so}^{i}(\mu_{0})$ to denote conductance near the $i = u~(l)$ for upper (lower) edge with bath chemical potential $\mu_{0}$. Further, the quantity $\tilde{\sigma}_{so}(\mu_{0}) = (\sigma_{so}^{u}(\mu_{0}) - \sigma_{so}^{l}(\mu_{0}))/2$ approaches quantization and gives the information of the number of edge modes crossing the chemical potential $\mu_{0}$.}
\label{tab:summed_conductance}
\end{table*}

\section{Summary}
\label{sec:4}

In summary, we studied the quench dynamics of a periodically driven graphene nanoribbon in both isolated and open-system settings. For the isolated (closed) graphene nanoribbon, we find that a sudden switch-on of the periodic drive (circularly polarized light) generates a persistent dc current about which the total bond current averaged over one driving period oscillates. The presence of this finite dc current serves as a signature of the topological phase of the driven graphene system. We further investigated the quench dynamics when the graphene nanoribbon is coupled to a featureless fermionic bath through a single coupling scale, $\Gamma$. In this case, both the current and the conductance increase with time and eventually saturate after a characteristic timescale. The saturation values are significantly larger in the topological phase than in the trivial phase. In both phases, the saturation indicates the emergence of a steady state. We find that the characteristic time required to reach the steady state is inversely proportional to the system-bath coupling strength, i.e., it scales as $1/\Gamma$.
To gain further insight into the understanding of saturation in the topological phase, we consider the limit $\Gamma \rightarrow 0$. In this limit, the featureless fermionic bath acts as an ideal thermodynamic reservoir, and the system requires an infinitely long time of $\mathcal{O}(1/\Gamma)$ to reach the steady state. In the steady state, the quasienergy bands acquire a time-independent staircase occupation profile described by Eq.~\eqref{eq:B3}. We find that the summed bond conductance along a given edge of the nanoribbon provides information about the chirality of the corresponding edge modes. Furthermore, the difference between the summed bond conductances near the upper and lower edges, divided by two, directly yields the number of chiral edge modes. We also find independently that the bond conductance is quantized after using the sum rule when calculated within the framework of two-terminal transport, as discussed in Appendix~\ref{app:d}.  

Our study opens several directions for future investigation. An important question is how long the topological signatures observed following the quench persist in the presence of electron-electron interactions. Since interactions can induce heating in periodically driven systems and eventually lead to thermalization toward an infinite-temperature state~\cite{PhysRevX.4.041048}, it is natural to ask on what timescale the topological features identified in this work remain robust. Another interesting direction would be the open system setup. In particular, it would be worthwhile to investigate how interactions modify the characteristic timescale required to reach the steady state, as well as the saturation values of the current and conductance. In general, understanding the interplay between periodic driving, interactions, and dissipation may provide deeper insights into topological phases of 
out-of-equilibrium quantum systems.

Our theoretical study can be tested in various experiments. The projected and staircase occupation of quasienergy bands can be observed in time-resolved angle-resolved spectroscopy. For experimental relevance, we note that for graphene $t_0\approx 2.7$ eV, so the high-frequency results at $\Omega=8t_0$ correspond to $\sim 22$ eV, which lies in the extreme ultraviolet range. The low-frequency data ($\Omega\sim 1$--$2\,t_0$) lie in the THz to mid-infrared range and are far more accessible with current ultrafast laser technology~\cite{doi:10.1126/science.1239834}. Further, the current following quench dynamics in a closed system can be detected using a SQUID magnetometer that responds to magnetization that occurs due to a local current~\cite{PhysRevLett.102.136802, shibata2015imaging}. Further our results in the presence of an ideal thermodynamic bath limit can be measured via a multiterminal transport study~\cite{PhysRevLett.113.266801, PhysRevB.96.165443}.

\section*{Acknowledgments}
A.D. acknowledges support from the National Post Doctoral Fellowship (NPDF) of the Anusandhan National Research Foundation (ANRF), India, under Grant No.~PDF/2025/001247, IISER TIRUPATI. And A.K. acknowledges funding support from ANRF via grant numbers ANRF/ARG/2025/002460/PS and ANRF/ARGM/2025/002682/TS, H.P acknowledge support from IIT Kanpur.

\appendix

\section{Derivation of current operator }
\label{app:a}

Here we discuss the current operator for a general periodically driven tight-binding model.
Consider a general tight-binding Hamiltonian
\begin{equation}
\begin{aligned}
\mathcal{H} = \sum_{i,j} t_{ij}\mathcal{C}_{i}^{\dagger}\mathcal{C}_{j},
\end{aligned}
\end{equation}
where \(t_{ij}\) is the hopping between sites $i$ and $j$. \(\mathcal{C}_{i}\) is the fermionic annihilation operator at site $i$.
In the presence of the periodic drive (say circularly polarized light), the hopping integral is modified following the Peierls substitution, \(t_{ij} \to t_{ij}e^{i\bm \delta_{ij}\cdot \bm A(t)}\). In the Peierls substitution, \(\bm \delta_{ij}\) is the displacement vector between sites $i$ and $j$, and \(\bm A(t) = (\cos(\Omega t), \sin(\Omega t)\) is the vector potential for the circularly polarized light. In the presence of the drive, the Hamiltonian is given by 
\begin{equation}
\begin{aligned}
\mathcal{H}(t) = \frac{1}{2}\sum_{ij} (t_{ij}e^{i\bm \delta_{ij}\cdot \bm A(t)}\mathcal{C}_{i}^{\dagger}\mathcal{C}_{j} + t_{ij}^{*}e^{-i\bm \delta_{ij}\cdot \bm A(t)}\mathcal{C}_{j}^{\dagger}\mathcal{C}_{i}).
\end{aligned}
\end{equation}
Now, in order to derive the current operator, we apply a static uniform weak electric field (the probe field) in the x direction. The vector potential for the probe field is given by \(\bm A_{pr} = (a_{pr},0)\). The probe field also enters the Hamiltonian through the Peierls substitution as
\begin{equation}
\begin{aligned}
\mathcal{H}(t) &= \frac{1}{2}\sum_{ij}~ ( t_{ij}e^{i\bm \delta_{ij}\cdot \bm A(t)}e^{i\delta^{x}_{ij}a_{pr}}\mathcal{C}_{i}^{\dagger}\mathcal{C}_{j} \\
&~~~~~~~~~~~~ + t_{ij}^{*}e^{-i\bm \delta_{ij}\cdot \bm A(t)}e^{-i\delta^{x}_{ij}a_{pr}}\mathcal{C}_{j}^{\dagger}\mathcal{C}_{i}).
\end{aligned}
\end{equation}
The $x$ component of the current operator is given by
\begin{equation}
\begin{aligned}
\mathcal{J}^{x} &= \frac{\partial \mathcal{H}}{\partial a_{pr}}\vert_{a_{pr} \to 0}\\
&= \frac{i}{2}\sum_{ij}\delta_{ij}^{x}\left(t_{ij}e^{i\bm \delta_{ij}\cdot \bm A(t)}\mathcal{C}_{i}^{\dagger}\mathcal{C}_{j} - \rm H.c. \right).
\end{aligned}
\end{equation}
Similarly, one can find the y-component of the current operator by applying the probe field in the 
y-direction. 

\section{Derivation of dc current }
\label{app:b}

In this section, we discuss the current averaged over one driving period as a function of stroboscopic time following a quantum quench. To characterize the quench dynamics, we define the average current over one driving period between two consecutive stroboscopic times as 
\begin{equation}
\begin{aligned}
\Bar{\mathcal{J}}_{A.V}(NT) = \frac{1}{T}\int_{N T}^{(N + 1) T} \bra{\psi_{g}}\mathcal{U}^{\dagger}(t)\mathcal{J}(t)\mathcal{U}(t)\ket{\psi_{g}} dt,
\end{aligned}
\end{equation}
where $|\psi_{g}\rangle$ is the many particle ground state of static system (half filled state).
Substitute \(t = NT + t'\) into the above equation, we 
obtain
\begin{equation}
\begin{aligned}
\Bar{\mathcal{J}}_{A.V}(NT) &= \frac{1}{ T}\int_{0}^{ T} \bra{\psi_{g}}\mathcal{U}^{\dagger}(t' + NT)\mathcal{J}(t' + NT)\\
&~~~~~~~~~~~~~\times\mathcal{U}(t' + NT )\ket{\psi_{g}} 
dt'.
\label{APP:b2}
\end{aligned}
\end{equation}
In the above equation, 
\begin{equation}
\begin{aligned}
\mathcal{U}(t' + NT) &= \mathcal{U}(t' + NT,0) =  \mathcal{U}(t' + NT,NT) \mathcal{U}( NT, 0)\\
&= \mathcal{U}(t',0) \mathcal{U}( NT, 0) = \mathcal{U}(t') \mathcal{U}( NT).
\end{aligned}
\end{equation}
The current operator is periodic in time, hence\(\mathcal{J}(t' + NT) = \mathcal{J}(t')\).
Hence we can rewrite Eq.~\eqref{APP:b2} as
\begin{equation}
\begin{aligned}
\Bar{\mathcal{J}}_{A.V}(NT) &= \frac{1}{ T}\int_{0}^{ T} \bra{\psi_{g}}\mathcal{U}^{\dagger}(NT) \mathcal{U}^{\dagger}(t')\mathcal{J}(t')\\
&~~~~~~~~~~~~~\times\mathcal{U}(t' )\mathcal{U}(NT)\ket{\psi_{g}} dt'.
\label{APP:b4}
\end{aligned}
\end{equation}
At zero temperature, for a noninteracting system, the expectation value of any  operator with respect to the many-particle ground state can be expressed as the sum of the expectation values of the operator over the occupied single-particle eigenstate. Thus, we can write the above expression as :
\begin{equation}
\begin{aligned}
\Bar{\mathcal{J}}_{A.V}(NT) &= \sum_{m \in occ}\frac{1}{ T}\int_{0}^{ T} \bra{m}\mathcal{U}^{\dagger}(NT) \mathcal{U}^{\dagger}(t')\mathcal{J}(t')\\
&~~~~~~~~~~~~~\times\mathcal{U}(t' )\mathcal{U}(NT)\ket{m} dt',
\label{APP:b5}
\end{aligned}
\end{equation}
where the symbol ``occ'' denotes the occupied states, while $\ket{m}$ denotes a single-particle state, i.e., an eigenstate of the static Hamiltonian. We can rewrite the sum over the occupied single-particle states as a sum over all available single-particle states by introducing the Fermi Dirac distribution $f(E_m)$ in the above equation, where $E_m$ is the corresponding eigenvalue of the static Hamiltonian. For $\mu=0$ and zero temperature, the Fermi Dirac distribution satisfies $f(E_m)=1$ for $E_m\leq\mu=0$ and $f(E_m)=0$ for $E_m>\mu$. Thus, the above expression can be written as:
\begin{equation}
\begin{aligned}
\Bar{\mathcal{J}}_{A.V}(NT) &= \sum_{m}f(E_{m})\frac{1}{ T}\int_{0}^{ T} \bra{m}\mathcal{U}^{\dagger}(NT) \mathcal{U}^{\dagger}(t')\mathcal{J}(t')\\
&\times\mathcal{U}(t' )\mathcal{U}(NT)\ket{m} dt'.
\label{APP:b6}
\end{aligned}
\end{equation}
We can further simplify the above expression for the average current by rewriting Eq.~\eqref{APP:b6} as,
\begin{equation}
\begin{aligned}
\Bar{\mathcal{J}}_{A.V}(NT) = \frac{1}{ T}\int_{0}^{T} \mathrm{Tr}\left[\rho^{th}\mathcal{U}^{\dagger}( N t')\mathcal{J}(N t')\mathcal{U}(N t')\right] dt'.
\end{aligned}
\end{equation}
In the above equation, \(\rho^{th} = \sum_{n} f(E_{n})\ket{n}\bra{n}\) is the thermal density matrix at zero temperature. After plugging in the zero temperature thermal density matrix expression and evolution operator into the above equation, the average current can be written as 
\begin{equation}
\begin{aligned}
\Bar{\mathcal{J}}_{A.V} &= \frac{1}{ T}\int_{0}^{T} \mathrm{Tr}\sum_{n, \alpha, \beta} e^{i N t'(\varepsilon_{\alpha} - \varepsilon_{\beta})} f(E_{n})\ket{n}\bra{n}\ket{\Phi_{\alpha}(0)}\\
&~~~~~~~~~~\times \bra{\Phi_{\alpha}(N t')}
\mathcal{J}(N t')\ket{u_{\beta}(N t')}\bra{\Phi_{\beta}(0)} dt'.
\end{aligned}
\end{equation}
In order to compute the dc value around which the average current oscillates, we take the large $N$ limit. At large \(N\), we need to keep only the \(\alpha = \beta \) terms. This is because the \(\alpha \ne \beta\) terms oscillate rapidly for large $N$ with different frequencies for different values of the momentum \(k_{x}\) in the case of graphene nanoribbon; hence summing over \(k_{x}\) lead to vanishing of the \(\alpha \ne \beta\) terms. In the light of the above statement and
after a little algebra, the above equation simplifies to, 
\begin{equation}
\begin{aligned}
\Bar{\mathcal{J}}_{dc} &= \frac{1}{ T}\int_{0}^{T} \sum_{n, \alpha, k_{x}}  f(E_{n})| \braket{n |\Phi_{\alpha}(0)}|^{2} \\
&~~~~~~~~~~~~~ \times \bra{\Phi_{\alpha}(N t')}
\mathcal{J}(N t')\ket{\Phi_{\alpha}(N t')} dt'.
\end{aligned}
\end{equation}
Thus, in a compact form the dc current is given by 
\begin{equation}
\begin{aligned}
 \Bar{\mathcal{J}}_{dc} = \sum_{\alpha, k_{x}} \rho_{\alpha}^{pr}(k_{x}) \Bar{\mathcal{J}_{\alpha}}(k_{x}), 
\end{aligned}
\end{equation}
where $\rho_{\alpha}^{pr} = \sum_{n} f(E_{n}) \braket{n | \Phi_{\alpha}(0)}|^{2}$ is the projected occupation of $\alpha$-th Floquet band, and $\Bar{\mathcal{J}}_{\alpha}  = \frac{1}{ T}\int_{0}^{T} \bra{\Phi_{\alpha}( t')} \mathcal{J}(t')\ket{\Phi_{\alpha}( t')} dt'$ is the current in the $\alpha$-th Floquet band

\begin{figure*}[t]
\centering
\includegraphics[width=1.0\textwidth]{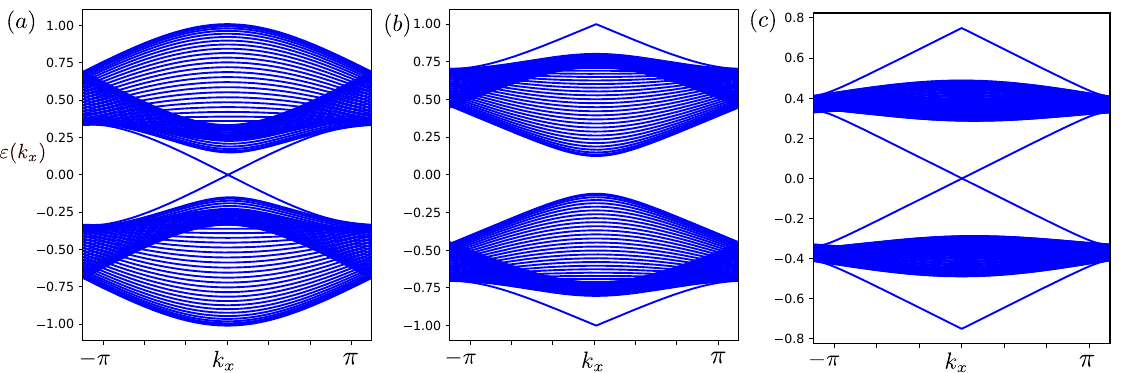}
\vspace{0.3cm}
\caption{(Color online) Quasienergy band structure of  graphene driven with a three-step driving protocol mentioned in Appendix~\ref{app:d}. In panel (a) $T = 2\pi/(3J)$ and $\Delta = 0$, in panel (b) $T= 2\pi/(2J)$ and $\Delta=1.3J$, and in panel (c) $T= 2\pi/(1.5J)$ and $\Delta=0.1J$.}  
\label{fig:APP-FIG-1}
\end{figure*}

\begin{figure*}[t]
\centering
\includegraphics[width=1.0\textwidth]{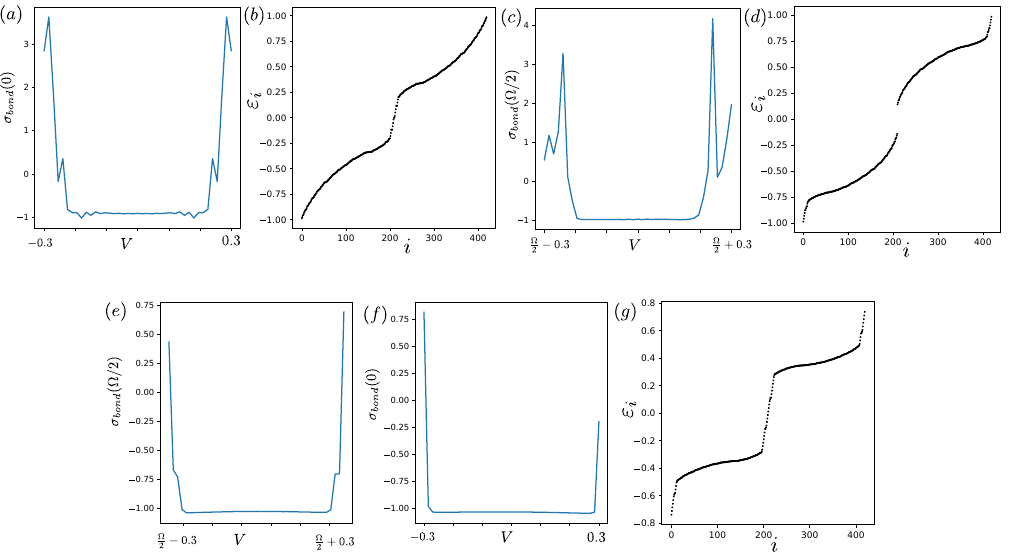}
\vspace{0.3cm}
\caption{(Color online)
Quantized bond conductance of periodically driven graphene under a three-step driving protocol after applying the sum rule and summing over a few bonds near the edge. Panels (a), (c), (e) and (f) show the bond conductance for different driving parameters, while panels (b), (d), and (g) display the corresponding quasienergy spectra. In panels (a) and (f), the bath chemical potential is set to zero, whereas in panels (c) and (e) it is placed at the Floquet zone boundary. For panels (a) and (b), the parameters are $T = 2\pi/3J$ and $\Delta = 0J$. Panels (c) and (d) correspond to $T = 2\pi/2J$ and $\Delta = 1.3J$. Panels (e) and (f) are obtained for $T = 2\pi/1.5J$ and $\Delta = 0.1J$. The quasienergy spectra shown in panels (b), (d), and (g) are plotted in the site basis for the same parameters used in the corresponding bond-conductance panels. Note that in the plots, $V$ is the bath chemical potential, i.e., $V = \mu$.
}
\label{fig:APP-FIG-2}
\end{figure*}

\section{Derivation of average bond conductance}
\label{app:c}

In this section, we discuss the computation of bond conductance in a regime in which the bath behaves as an ideal thermodynamic bath. When the bath behaves as a featureless fermionic bath and there is a single coupling scale \(\Gamma\) between the system and the bath, the effective density matrix of the system as a function of the bath chemical potential is given Eq.~\eqref{eq:B1} of the main text. The conductance is given by
\begin{equation}
\begin{aligned}
\sigma_{A.V}(\mu, NT) = \frac{2\pi}{T} \int_{N T}^{(N+1)T} \mathrm{Tr} \left[\frac{d\rho(t)}{d\mu}\mathcal{J}(t)\right] dt.
\end{aligned}
\end{equation}
In the above equation, the density matrix $\rho(t)$ is given by Eq.~\eqref{eq:B2} of the main text. With a little algebra one can show that at zero temperature, the conductance is given by
\begin{eqnarray}
&& \sigma_{A.V}(\mu , NT) \nonumber \\
&& = \frac{2\pi}{T} \int_{N T}^{(N+1)T} \mathrm{Tr} \left[\mathcal{U}_{\Gamma}(t, \mu)\mathcal{J}(t)\mathcal{U}^{\dagger}_{\Gamma}(t, \mu)\right] dt.
\end{eqnarray}
In the limit of \(\Gamma \to 0\), the bath behaves as an ideal thermodynamic bath, and the effective density matrix of system is given by Eq.~\eqref{eq:B3} of the main text,
\begin{equation}
\begin{aligned}
\lim_{\Gamma \to 0} ~\rho_{s}(\bm k, t) &= \sum_{\alpha} p_{\alpha}(\bm k)\ket{\Psi_{\alpha}(\bm k, t)}\bra{\Psi_{\alpha}(\bm k, t)}, \\
p_{\alpha}(\bm k) &= \sum_{l=-\infty}^{\infty} f_{0}(\varepsilon_{\bm k,\alpha}+l\Omega - \mu_{0})
\langle \Phi_{\alpha}^{l} | \Phi_{\alpha}^{l} \rangle .
\end{aligned}
\end{equation}
In this regime, the system takes a long time of \(\mathcal{O}(1/\Gamma)\) to reach the steady state.
The average conductance over one driving period with the above density matrix is given by
\begin{equation}
\sigma(\mu_{0})
= \frac{1}{2\pi}\int d \bm k ~\frac{1}{T} \int_{NT}^{(N+1)T}
\Tr\!\left[
\frac{d\rho_{s}(\bm k, t)}{d\mu_{0}}\, \mathcal{J}(t)
\right] dt.
\end{equation}
One can sum the conductance for all value of bath chemical potential shifted by integer multiples of the driving frequency, i.e.,
\begin{equation}
\begin{aligned}
 \sigma_{ss}(\mu_{0}) &= \sum_{m}\sigma(\mu_{0} + m\Omega )\\
&= \sum_{m}\frac{1}{2\pi}\int d \bm k ~\frac{1}{T} \int_{NT}^{(N+1)T}
\Tr\!\left[
\frac{d\rho_{s}(\bm k, t)}{d\mu_{0}}\, \mathcal{J}(t)
\right] dt.   
\end{aligned}
\end{equation}
Upon substituting the above density matrix, we obtain
\begin{equation}
\begin{aligned}
 \sigma_{ss}(\mu_{0})
&= \frac{1}{2\pi}\int d\bm k ~\frac{1}{T} \int_{NT}^{(N+1)T}\\
&\times \Tr\!\left[
\sum_{\alpha} \frac{d\tilde{p}_{\alpha}(\bm k)}{d\mu_{0}}
\ket{\Phi_{\alpha}(\bm k, t)}\bra{\Phi_{\alpha}(\bm k, t)} \mathcal{J}(t)
\right] dt.
\end{aligned}
\end{equation}
In the above expression,
\begin{equation}
\frac{d\tilde{p}_{\alpha}(\bm k)}{d\mu_{0}} = \sum_{m=-\infty}^{\infty}\sum_{l=-\infty}^{\infty} \frac{df_{0}(\varepsilon_{\bm k,\alpha}+l\Omega - \mu_{0} - m\Omega)}{d\mu_{0}}
\langle \Phi_{\alpha}^{l} | \Phi_{\alpha}^{l} \rangle.
\end{equation}
At zero temperature, the derivative can be written as
\begin{align}
\frac{d\tilde{p}_{\alpha}(\bm k)}{d\mu_{0}}
&= \sum_{m=-\infty}^{\infty}\sum_{l=-\infty}^{\infty} \delta(\varepsilon_{\bm k,\alpha}+l\Omega - \mu_{0} - m\Omega)
\langle \Phi_{\alpha}^{l} | \Phi_{\alpha}^{l} \rangle.
\end{align}
With this, conductance can be rewritten as,
\begin{equation}
\begin{aligned}
& \sigma_{ss}(\mu_{0})
= \frac{1}{2\pi}\int d\bm k ~\sum_{\alpha} \frac{d\tilde{p}_{\alpha}(\bm k)}{d\mu_{0}}
\frac{1}{T} \int_{NT}^{(N+1)T}\\
&\times\sum_{\beta}
\bra{\Phi_{\beta}(\bm k, t)} \Phi_{\alpha}(\bm k, t)\rangle
\bra{\Phi_{\alpha}(\bm k, t)} \mathcal{J}(t) \ket{\Phi_{\beta}(\bm k, t)} dt. 
\end{aligned}
\end{equation}
Using orthogonality, we can write,
\begin{equation}
\begin{aligned}
  \sigma_{ss}(\mu_{0})
&= \frac{1}{2\pi}\int d\bm k \sum_{\alpha} \frac{d\tilde{p}_{\alpha}(\bm k)}{d\mu_{0}}\\
&~~~~~~ \times
\frac{1}{T} \int_{NT}^{(N+1)T}
\bra{\Phi_{\alpha}(\bm k, t)} \mathcal{J}(t) \ket{\Phi_{\alpha}(\bm k, t)} dt.  
\end{aligned}
\end{equation}
Due to periodicity, we then obtain
\begin{equation}
\begin{aligned}
 \sigma_{ss}(\mu_{0})
&= \frac{1}{2\pi}\int d \bm k \sum_{\alpha} \frac{d\tilde{p}_{\alpha}(\bm k)}{d\mu_{0}}\\
&\times\frac{1}{T} \int_{0}^{T}
\bra{\Phi_{\alpha}(\bm k, t)} \mathcal{J}(t) \ket{\Phi_{\alpha}(\bm k, t)} dt.
\end{aligned}
\end{equation}

The conductance then becomes
\begin{equation}
\begin{aligned}
\sigma_{ss}(\mu_{0})
&= \frac{1}{2\pi}\int d\bm k \sum_{\alpha} \sum_{\ell,m=-\infty}^{\infty}
\delta\!\left(\varepsilon_{\bm k,\alpha} + \ell\Omega - \mu_{0} - m\Omega\right)
\\
&~~~~ \times \langle \Phi_{\alpha}^{l} | \Phi_{\alpha}^{l} \rangle
~\frac{1}{T} \int_{0}^{T}
\bra{\Phi_{\alpha}(\bm k, t)} J(t) \ket{\Phi_{\alpha}(\bm k, t)} dt.
\end{aligned}
\end{equation}

\section{Two-terminal transport calculation of bond conductance}
\label{app:d}

In this section, we demonstrate the quantization of the bond conductance in a two-terminal geometry by imposing the sum rule and summing the conductance over several bonds near the edge. For clarity, we focus on a three-step driving protocol, as it yields a sufficiently large bulk gap, enabling reliable transport calculations in the two-terminal setup.

Following Ref.~\cite{PhysRevB.101.041403}, the bond current is given by
\begin{equation}
\begin{aligned}
j_{ij} = \frac{1}{T}\int_{0}^{T} 2 ~\mathrm{Im}\left[J_{ij}(t)\langle \mathrm{C}_{i}^{\dagger}(t)\mathrm{C}_{j}(t)\rangle\right] dt,
\label{eq:app-d1}
\end{aligned} 
\end{equation}
where $\mathcal{C}_{i}$ is the annihilation operator for electrons at site $i$, and $J_{ij}(t)$ is the current at time $t$ between sites $i$ and $j$. 
In terms of the Floquet Green's function,
the electron operator is given by
\begin{equation}
\begin{aligned}
\mathcal{C}_{i}(t) = \sum_{\lambda, k_{1}} \int \frac{d\omega_{1}}{2\pi}e^{-i\omega_{1} t}e^{-i k_{1} \omega_{1} t}G^{k_{1}}(\omega_{1})_{ii_{1}}h_{i_{1}}^{\lambda_{1}}(\omega_{1}),
\label{eq:app-d2}
\end{aligned}
\end{equation}
where $\lambda = (L,R)$ denotes the left and right leads. The fluctuation-dissipation relation satisfied by noise term $h^{\lambda}(\omega)$ is given by~\cite{PhysRevB.101.041403}
\begin{equation}
\begin{aligned}
&\langle h^{\lambda_{1}\dagger}_{i_{1}}(\omega_{1}) h^{\lambda_{2}}_{j_{1}}(\omega{2})\rangle\\
&= \mathcal{K}^{\lambda_{1}}_{\alpha i_{1}}\mathcal{K}^{\lambda_{2}*}_{\beta j_{1}}(2\pi)^{2}\delta_{\lambda_{1},\lambda_{2}}\rho(\omega)_{i_{i}j_{1}}^{\lambda_{1}}f^{\lambda_{1}}(\omega_{1})\delta(\omega_{1} -\omega_{2}),
\label{eq:app-d3}
\end{aligned}
\end{equation}
where $\rho^{\lambda}(\omega)$ is the density matrix of the lead. In the wide-band limit, this is constant and 
does not depend upon $\omega$. The matrix $\mathcal{K}$ is a coupling matrix for the system coupled to a bath, and $f^{\lambda}(\omega)$ is the Fermi-Dirac distribution function for the lead.
Upon plugging Eq.~\eqref{eq:app-d3} and Eq.~\eqref{eq:app-d2} into Eq.~\eqref{eq:app-d1} and after some simplification, we obtain
\begin{equation}
\begin{aligned}
j_{ij} &= \sum_{\lambda k_{1} k_{2}} \frac{1}{T}\int_{0}^{T} dt J_{ij}(t) \\
&\times e^{i\Omega(k_{1} - k_{2})t}\int d\omega \left(G^{k_{1}\dagger}(\omega)V^{\lambda}G^{k_{2}}\right)_{ij}f^{\lambda}(\omega),
\end{aligned}
\end{equation}
where the matrix $V^{\lambda} = \mathcal{K}^{\dagger}\rho\mathcal{K}$.
In order to compute the bond conductance we assume that the
left lead chemical potential is at $\mu$, i.e., $f^{\lambda = L}(\omega) = f(\omega - \mu)$, and the right lead is at a chemical potential $\mu + \delta\mu$, i.e., $f^{\lambda = R}(\omega) = f(\omega - \mu - \delta\mu)$. The bond conductance units of $e^{2}/h$ is given by
\begin{equation}
\begin{aligned}
\sigma_{ij}(\mu)  &= \frac{e}{\hbar}\frac{dj_{ij}}{d\delta\mu}|_{\delta\mu \to 0}\\
&= \sum_{ k_{1} k_{2}} \frac{2\pi}{T}\int_{0}^{T} dt
J_{ij}(t) e^{i\Omega(k_{1} - k_{2})t}\\
 &\times\int d\omega \left(G^{k_{1}\dagger}(\mu)V^{L}G^{k_{2}}\right)_{ij} + \left(G^{k_{1}\dagger}(\mu)V^{R}G^{k_{2}}\right)_{ij}.
\end{aligned}
\end{equation}

 We consider a three-step drive protocol. Following Ref.~\cite{PhysRevResearch.1.022003}, the Hamiltonian for such a drive is given by
\begin{eqnarray}
\mathcal{H}_{th}(\bm k ,t) &=& - ~\sum_{i=1}^{3} J_i(t)
\Big[ \cos(\bm k\!\cdot\bm \delta_i)\,\sigma_x +
\sin(\bm k\!\cdot \bm \delta_i)\,\sigma_y \Big] \nonumber \\
&& + ~\frac{\Delta}{2}\,\sigma_z.
\end{eqnarray}
In the above equation, $J_{i}(t) = J$ is the 
nearest-neighbor hopping along the three 
nearest-neighbor bonds $\bm \delta_{i}$. In the 
three-step driving protocol, the driving period $T$ is divided into three periods. During $i$-th period for \(i = {1,2,3}\), hopping in only the $i$-th direction is allowed, and the other two hoppings are not allowed. This driving protocol breaks the 
time-reversal symmetry and also leads to the appearance of edge modes and anomalous edge modes. The armchair nanoribbon band structure for this driving protocol is 
shown in Fig.~\ref{fig:APP-FIG-1}. 

Using the sum rule and after adding a few sites parallel to the edge of the nanoribbon along the perpendicular direction of the edge, the bond conductance is given by
\begin{equation}
\begin{aligned}
\sigma_{bond}(\mu) = \sum_{i = 0}^{i = N_{b}} \sum_{p = -M}^{M}\sigma_{4i \to 4i+1}(\mu + p\Omega),
\end{aligned}
\end{equation}
where $M$ is the number of Floquet copies taken into account. For our numerics we take $M = 5$.
We refer to Fig.~\ref{fig:APP-FIG-2} for the conductance for various values of the driving frequency and the mass term.

\bibliography{references}

\begin{thebibliography}{80}%
\makeatletter
\providecommand \@ifxundefined [1]{%
 \@ifx{#1\undefined}
}%
\providecommand \@ifnum [1]{%
 \ifnum #1\expandafter \@firstoftwo
 \else \expandafter \@secondoftwo
 \fi
}%
\providecommand \@ifx [1]{%
 \ifx #1\expandafter \@firstoftwo
 \else \expandafter \@secondoftwo
 \fi
}%
\providecommand \natexlab [1]{#1}%
\providecommand \enquote  [1]{``#1''}%
\providecommand \bibnamefont  [1]{#1}%
\providecommand \bibfnamefont [1]{#1}%
\providecommand \citenamefont [1]{#1}%
\providecommand \href@noop [0]{\@secondoftwo}%
\providecommand \href [0]{\begingroup \@sanitize@url \@href}%
\providecommand \@href[1]{\@@startlink{#1}\@@href}%
\providecommand \@@href[1]{\endgroup#1\@@endlink}%
\providecommand \@sanitize@url [0]{\catcode `\\12\catcode `\$12\catcode `\&12\catcode `\#12\catcode `\^12\catcode `\_12\catcode `\%12\relax}%
\providecommand \@@startlink[1]{}%
\providecommand \@@endlink[0]{}%
\providecommand \url  [0]{\begingroup\@sanitize@url \@url }%
\providecommand \@url [1]{\endgroup\@href {#1}{\urlprefix }}%
\providecommand \urlprefix  [0]{URL }%
\providecommand \Eprint [0]{\href }%
\providecommand \doibase [0]{https://doi.org/}%
\providecommand \selectlanguage [0]{\@gobble}%
\providecommand \bibinfo  [0]{\@secondoftwo}%
\providecommand \bibfield  [0]{\@secondoftwo}%
\providecommand \translation [1]{[#1]}%
\providecommand \BibitemOpen [0]{}%
\providecommand \bibitemStop [0]{}%
\providecommand \bibitemNoStop [0]{.\EOS\space}%
\providecommand \EOS [0]{\spacefactor3000\relax}%
\providecommand \BibitemShut  [1]{\csname bibitem#1\endcsname}%
\let\auto@bib@innerbib\@empty
%</preamble>
\bibitem [{\citenamefont {Wen}(2017)}]{RevModPhys.89.041004}%
  \BibitemOpen
  \bibfield  {author} {\bibinfo {author} {\bibfnamefont {X.-G.}\ \bibnamefont {Wen}},\ }\bibfield  {title} {\bibinfo {title} {Colloquium: Zoo of quantum-topological phases of matter},\ }\href {https://doi.org/10.1103/RevModPhys.89.041004} {\bibfield  {journal} {\bibinfo  {journal} {Rev. Mod. Phys.}\ }\textbf {\bibinfo {volume} {89}},\ \bibinfo {pages} {041004} (\bibinfo {year} {2017})}\BibitemShut {NoStop}%
\bibitem [{\citenamefont {Kane}\ and\ \citenamefont {Mele}(2005)}]{PhysRevLett.95.226801}%
  \BibitemOpen
  \bibfield  {author} {\bibinfo {author} {\bibfnamefont {C.~L.}\ \bibnamefont {Kane}}\ and\ \bibinfo {author} {\bibfnamefont {E.~J.}\ \bibnamefont {Mele}},\ }\bibfield  {title} {\bibinfo {title} {Quantum spin {H}all effect in graphene},\ }\href {https://doi.org/10.1103/PhysRevLett.95.226801} {\bibfield  {journal} {\bibinfo  {journal} {Phys. Rev. Lett.}\ }\textbf {\bibinfo {volume} {95}},\ \bibinfo {pages} {226801} (\bibinfo {year} {2005})}\BibitemShut {NoStop}%
\bibitem [{\citenamefont {Fu}\ \emph {et~al.}(2007)\citenamefont {Fu}, \citenamefont {Kane},\ and\ \citenamefont {Mele}}]{PhysRevLett.98.106803}%
  \BibitemOpen
  \bibfield  {author} {\bibinfo {author} {\bibfnamefont {L.}~\bibnamefont {Fu}}, \bibinfo {author} {\bibfnamefont {C.~L.}\ \bibnamefont {Kane}},\ and\ \bibinfo {author} {\bibfnamefont {E.~J.}\ \bibnamefont {Mele}},\ }\bibfield  {title} {\bibinfo {title} {Topological insulators in three dimensions},\ }\href {https://doi.org/10.1103/PhysRevLett.98.106803} {\bibfield  {journal} {\bibinfo  {journal} {Phys. Rev. Lett.}\ }\textbf {\bibinfo {volume} {98}},\ \bibinfo {pages} {106803} (\bibinfo {year} {2007})}\BibitemShut {NoStop}%
\bibitem [{\citenamefont {Wan}\ \emph {et~al.}(2011)\citenamefont {Wan}, \citenamefont {Turner}, \citenamefont {Vishwanath},\ and\ \citenamefont {Savrasov}}]{PhysRevB.83.205101}%
  \BibitemOpen
  \bibfield  {author} {\bibinfo {author} {\bibfnamefont {X.}~\bibnamefont {Wan}}, \bibinfo {author} {\bibfnamefont {A.~M.}\ \bibnamefont {Turner}}, \bibinfo {author} {\bibfnamefont {A.}~\bibnamefont {Vishwanath}},\ and\ \bibinfo {author} {\bibfnamefont {S.~Y.}\ \bibnamefont {Savrasov}},\ }\bibfield  {title} {\bibinfo {title} {Topological semimetal and {F}ermi-arc surface states in the electronic structure of pyrochlore iridates},\ }\href {https://doi.org/10.1103/PhysRevB.83.205101} {\bibfield  {journal} {\bibinfo  {journal} {Phys. Rev. B}\ }\textbf {\bibinfo {volume} {83}},\ \bibinfo {pages} {205101} (\bibinfo {year} {2011})}\BibitemShut {NoStop}%
\bibitem [{\citenamefont {Kitaev}(2001)}]{kitaev2001unpaired}%
  \BibitemOpen
  \bibfield  {author} {\bibinfo {author} {\bibfnamefont {A.~Y.}\ \bibnamefont {Kitaev}},\ }\bibfield  {title} {\bibinfo {title} {Unpaired {M}ajorana fermions in quantum wires},\ }\href@noop {} {\bibfield  {journal} {\bibinfo  {journal} {Physics-uspekhi}\ }\textbf {\bibinfo {volume} {44}},\ \bibinfo {pages} {131} (\bibinfo {year} {2001})}\BibitemShut {NoStop}%
\bibitem [{\citenamefont {Lahtinen}\ and\ \citenamefont {Pachos}(2017{\natexlab{a}})}]{Lahtinen2017}%
  \BibitemOpen
  \bibfield  {author} {\bibinfo {author} {\bibfnamefont {V.}~\bibnamefont {Lahtinen}}\ and\ \bibinfo {author} {\bibfnamefont {J.~K.}\ \bibnamefont {Pachos}},\ }\bibinfo {title} {Topological aspects of quantum information processing},\ in\ \href {https://doi.org/10.1007/978-981-10-6841-6_18} {\emph {\bibinfo {booktitle} {Topology and Condensed Matter Physics}}},\ \bibinfo {editor} {edited by\ \bibinfo {editor} {\bibfnamefont {S.~M.}\ \bibnamefont {Bhattacharjee}}, \bibinfo {editor} {\bibfnamefont {M.}~\bibnamefont {Mj}},\ and\ \bibinfo {editor} {\bibfnamefont {A.}~\bibnamefont {Bandyopadhyay}}}\ (\bibinfo  {publisher} {Springer Singapore},\ \bibinfo {address} {Singapore},\ \bibinfo {year} {2017})\ pp.\ \bibinfo {pages} {471--500}\BibitemShut {NoStop}%
\bibitem [{\citenamefont {Bonderson}\ and\ \citenamefont {Nayak}(2013)}]{PhysRevB.87.195451}%
  \BibitemOpen
  \bibfield  {author} {\bibinfo {author} {\bibfnamefont {P.}~\bibnamefont {Bonderson}}\ and\ \bibinfo {author} {\bibfnamefont {C.}~\bibnamefont {Nayak}},\ }\bibfield  {title} {\bibinfo {title} {Quasi-topological phases of matter and topological protection},\ }\href {https://doi.org/10.1103/PhysRevB.87.195451} {\bibfield  {journal} {\bibinfo  {journal} {Phys. Rev. B}\ }\textbf {\bibinfo {volume} {87}},\ \bibinfo {pages} {195451} (\bibinfo {year} {2013})}\BibitemShut {NoStop}%
\bibitem [{\citenamefont {Lahtinen}\ and\ \citenamefont {Pachos}(2017{\natexlab{b}})}]{10.21468/SciPostPhys.3.3.021}%
  \BibitemOpen
  \bibfield  {author} {\bibinfo {author} {\bibfnamefont {V.}~\bibnamefont {Lahtinen}}\ and\ \bibinfo {author} {\bibfnamefont {J.~K.}\ \bibnamefont {Pachos}},\ }\bibfield  {title} {\bibinfo {title} {{A Short Introduction to Topological Quantum Computation}},\ }\href {https://doi.org/10.21468/SciPostPhys.3.3.021} {\bibfield  {journal} {\bibinfo  {journal} {SciPost Phys.}\ }\textbf {\bibinfo {volume} {3}},\ \bibinfo {pages} {021} (\bibinfo {year} {2017}{\natexlab{b}})}\BibitemShut {NoStop}%
\bibitem [{\citenamefont {Cava}\ \emph {et~al.}(2013)\citenamefont {Cava}, \citenamefont {Ji}, \citenamefont {Fuccillo}, \citenamefont {Gibson},\ and\ \citenamefont {Hor}}]{C3TC30186A}%
  \BibitemOpen
  \bibfield  {author} {\bibinfo {author} {\bibfnamefont {R.~J.}\ \bibnamefont {Cava}}, \bibinfo {author} {\bibfnamefont {H.}~\bibnamefont {Ji}}, \bibinfo {author} {\bibfnamefont {M.~K.}\ \bibnamefont {Fuccillo}}, \bibinfo {author} {\bibfnamefont {Q.~D.}\ \bibnamefont {Gibson}},\ and\ \bibinfo {author} {\bibfnamefont {Y.~S.}\ \bibnamefont {Hor}},\ }\bibfield  {title} {\bibinfo {title} {Crystal structure and chemistry of topological insulators},\ }\href {https://doi.org/10.1039/C3TC30186A} {\bibfield  {journal} {\bibinfo  {journal} {J. Mater. Chem. C}\ }\textbf {\bibinfo {volume} {1}},\ \bibinfo {pages} {3176} (\bibinfo {year} {2013})}\BibitemShut {NoStop}%
\bibitem [{\citenamefont {Lian}\ \emph {et~al.}(2019)\citenamefont {Lian}, \citenamefont {Wang}, \citenamefont {Lu}, \citenamefont {Huang}, \citenamefont {Wang}, \citenamefont {Yuan}, \citenamefont {Zhang}, \citenamefont {Ouyang}, \citenamefont {Wang}, \citenamefont {Huang}, \citenamefont {He}, \citenamefont {Chang}, \citenamefont {Deng},\ and\ \citenamefont {Duan}}]{PhysRevLett.122.210503}%
  \BibitemOpen
  \bibfield  {author} {\bibinfo {author} {\bibfnamefont {W.}~\bibnamefont {Lian}}, \bibinfo {author} {\bibfnamefont {S.-T.}\ \bibnamefont {Wang}}, \bibinfo {author} {\bibfnamefont {S.}~\bibnamefont {Lu}}, \bibinfo {author} {\bibfnamefont {Y.}~\bibnamefont {Huang}}, \bibinfo {author} {\bibfnamefont {F.}~\bibnamefont {Wang}}, \bibinfo {author} {\bibfnamefont {X.}~\bibnamefont {Yuan}}, \bibinfo {author} {\bibfnamefont {W.}~\bibnamefont {Zhang}}, \bibinfo {author} {\bibfnamefont {X.}~\bibnamefont {Ouyang}}, \bibinfo {author} {\bibfnamefont {X.}~\bibnamefont {Wang}}, \bibinfo {author} {\bibfnamefont {X.}~\bibnamefont {Huang}}, \bibinfo {author} {\bibfnamefont {L.}~\bibnamefont {He}}, \bibinfo {author} {\bibfnamefont {X.}~\bibnamefont {Chang}}, \bibinfo {author} {\bibfnamefont {D.-L.}\ \bibnamefont {Deng}},\ and\ \bibinfo {author} {\bibfnamefont {L.}~\bibnamefont {Duan}},\ }\bibfield  {title} {\bibinfo {title} {Machine learning topological phases with a solid-state quantum simulator},\ }\href
  {https://doi.org/10.1103/PhysRevLett.122.210503} {\bibfield  {journal} {\bibinfo  {journal} {Phys. Rev. Lett.}\ }\textbf {\bibinfo {volume} {122}},\ \bibinfo {pages} {210503} (\bibinfo {year} {2019})}\BibitemShut {NoStop}%
\bibitem [{\citenamefont {Pickup}\ \emph {et~al.}(2020)\citenamefont {Pickup}, \citenamefont {Sigurdsson}, \citenamefont {Ruostekoski},\ and\ \citenamefont {Lagoudakis}}]{Pickup2020}%
  \BibitemOpen
  \bibfield  {author} {\bibinfo {author} {\bibfnamefont {L.}~\bibnamefont {Pickup}}, \bibinfo {author} {\bibfnamefont {H.}~\bibnamefont {Sigurdsson}}, \bibinfo {author} {\bibfnamefont {J.}~\bibnamefont {Ruostekoski}},\ and\ \bibinfo {author} {\bibfnamefont {P.~G.}\ \bibnamefont {Lagoudakis}},\ }\bibfield  {title} {\bibinfo {title} {Synthetic band-structure engineering in polariton crystals with non-{H}ermitian topological phases},\ }\href {https://doi.org/10.1038/s41467-020-18213-1} {\bibfield  {journal} {\bibinfo  {journal} {Nat. Commun.}\ }\textbf {\bibinfo {volume} {11}},\ \bibinfo {pages} {4431} (\bibinfo {year} {2020})}\BibitemShut {NoStop}%
\bibitem [{\citenamefont {Arg{\"u}ello-Luengo}\ \emph {et~al.}(2024)\citenamefont {Arg{\"u}ello-Luengo}, \citenamefont {Bhattacharya}, \citenamefont {Celi}, \citenamefont {Chhajlany}, \citenamefont {Grass}, \citenamefont {P{\l}odzie{\'n}}, \citenamefont {Rakshit}, \citenamefont {Salamon}, \citenamefont {Stornati}, \citenamefont {Tarruell},\ and\ \citenamefont {Lewenstein}}]{ArguelloLuengo2024}%
  \BibitemOpen
  \bibfield  {author} {\bibinfo {author} {\bibfnamefont {J.}~\bibnamefont {Arg{\"u}ello-Luengo}}, \bibinfo {author} {\bibfnamefont {U.}~\bibnamefont {Bhattacharya}}, \bibinfo {author} {\bibfnamefont {A.}~\bibnamefont {Celi}}, \bibinfo {author} {\bibfnamefont {R.~W.}\ \bibnamefont {Chhajlany}}, \bibinfo {author} {\bibfnamefont {T.}~\bibnamefont {Grass}}, \bibinfo {author} {\bibfnamefont {M.}~\bibnamefont {P{\l}odzie{\'n}}}, \bibinfo {author} {\bibfnamefont {D.}~\bibnamefont {Rakshit}}, \bibinfo {author} {\bibfnamefont {T.}~\bibnamefont {Salamon}}, \bibinfo {author} {\bibfnamefont {P.}~\bibnamefont {Stornati}}, \bibinfo {author} {\bibfnamefont {L.}~\bibnamefont {Tarruell}},\ and\ \bibinfo {author} {\bibfnamefont {M.}~\bibnamefont {Lewenstein}},\ }\bibfield  {title} {\bibinfo {title} {Synthetic dimensions for topological and quantum phases},\ }\href {https://doi.org/10.1038/s42005-024-01636-3} {\bibfield  {journal} {\bibinfo  {journal} {Commun. Phys.}\ }\textbf {\bibinfo {volume} {7}},\ \bibinfo {pages} {143}
  (\bibinfo {year} {2024})}\BibitemShut {NoStop}%
\bibitem [{\citenamefont {Mitra}(2018)}]{annurev:/content/journals/10.1146/annurev-conmatphys-031016-025451}%
  \BibitemOpen
  \bibfield  {author} {\bibinfo {author} {\bibfnamefont {A.}~\bibnamefont {Mitra}},\ }\bibfield  {title} {\bibinfo {title} {Quantum quench dynamics},\ }\href {https://doi.org/https://doi.org/10.1146/annurev-conmatphys-031016-025451} {\bibfield  {journal} {\bibinfo  {journal} {Annual Review of Condensed Matter Physics}\ }\textbf {\bibinfo {volume} {9}},\ \bibinfo {pages} {245} (\bibinfo {year} {2018})}\BibitemShut {NoStop}%
\bibitem [{\citenamefont {Eckardt}\ and\ \citenamefont {Anisimovas}(2015)}]{eckardt2015high}%
  \BibitemOpen
  \bibfield  {author} {\bibinfo {author} {\bibfnamefont {A.}~\bibnamefont {Eckardt}}\ and\ \bibinfo {author} {\bibfnamefont {E.}~\bibnamefont {Anisimovas}},\ }\bibfield  {title} {\bibinfo {title} {High-frequency approximation for periodically driven quantum systems from a {F}loquet-space perspective},\ }\href {https://doi.org/10.1088/1367-2630/17/9/093039} {\bibfield  {journal} {\bibinfo  {journal} {New Journal of physics}\ }\textbf {\bibinfo {volume} {17}},\ \bibinfo {pages} {093039} (\bibinfo {year} {2015})}\BibitemShut {NoStop}%
\bibitem [{\citenamefont {Bukov}\ \emph {et~al.}(2015)\citenamefont {Bukov}, \citenamefont {D'Alessio},\ and\ \citenamefont {Polkovnikov}}]{Bukov2015}%
  \BibitemOpen
  \bibfield  {author} {\bibinfo {author} {\bibfnamefont {M.}~\bibnamefont {Bukov}}, \bibinfo {author} {\bibfnamefont {L.}~\bibnamefont {D'Alessio}},\ and\ \bibinfo {author} {\bibfnamefont {A.}~\bibnamefont {Polkovnikov}},\ }\bibfield  {title} {\bibinfo {title} {Universal high-frequency behavior of periodically driven systems: from dynamical stabilization to {F}loquet engineering},\ }\href {https://doi.org/10.1080/00018732.2015.1055918} {\bibfield  {journal} {\bibinfo  {journal} {Advances in Physics}\ }\textbf {\bibinfo {volume} {64}},\ \bibinfo {pages} {139} (\bibinfo {year} {2015})}\BibitemShut {NoStop}%
\bibitem [{\citenamefont {Sen}\ \emph {et~al.}(2021)\citenamefont {Sen}, \citenamefont {Sen},\ and\ \citenamefont {Sengupta}}]{sen2021}%
  \BibitemOpen
  \bibfield  {author} {\bibinfo {author} {\bibfnamefont {A.}~\bibnamefont {Sen}}, \bibinfo {author} {\bibfnamefont {D.}~\bibnamefont {Sen}},\ and\ \bibinfo {author} {\bibfnamefont {K.}~\bibnamefont {Sengupta}},\ }\bibfield  {title} {\bibinfo {title} {Analytic approaches to periodically driven closed quantum systems: methods and applications},\ }\href {https://doi.org/10.1088/1361-648X/ac1b61} {\bibfield  {journal} {\bibinfo  {journal} {Journal of Physics: Condensed Matter}\ }\textbf {\bibinfo {volume} {33}},\ \bibinfo {pages} {443003} (\bibinfo {year} {2021})}\BibitemShut {NoStop}%
\bibitem [{\citenamefont {Oka}\ and\ \citenamefont {Aoki}(2009)}]{PhysRevB.79.081406}%
  \BibitemOpen
  \bibfield  {author} {\bibinfo {author} {\bibfnamefont {T.}~\bibnamefont {Oka}}\ and\ \bibinfo {author} {\bibfnamefont {H.}~\bibnamefont {Aoki}},\ }\bibfield  {title} {\bibinfo {title} {Photovoltaic {H}all effect in graphene},\ }\href {https://doi.org/10.1103/PhysRevB.79.081406} {\bibfield  {journal} {\bibinfo  {journal} {Phys. Rev. B}\ }\textbf {\bibinfo {volume} {79}},\ \bibinfo {pages} {081406(R)} (\bibinfo {year} {2009})}\BibitemShut {NoStop}%
\bibitem [{\citenamefont {Kundu}\ \emph {et~al.}(2014)\citenamefont {Kundu}, \citenamefont {Fertig},\ and\ \citenamefont {Seradjeh}}]{PhysRevLett.113.236803}%
  \BibitemOpen
  \bibfield  {author} {\bibinfo {author} {\bibfnamefont {A.}~\bibnamefont {Kundu}}, \bibinfo {author} {\bibfnamefont {H.~A.}\ \bibnamefont {Fertig}},\ and\ \bibinfo {author} {\bibfnamefont {B.}~\bibnamefont {Seradjeh}},\ }\bibfield  {title} {\bibinfo {title} {Effective theory of {F}loquet topological transitions},\ }\href {https://doi.org/10.1103/PhysRevLett.113.236803} {\bibfield  {journal} {\bibinfo  {journal} {Phys. Rev. Lett.}\ }\textbf {\bibinfo {volume} {113}},\ \bibinfo {pages} {236803} (\bibinfo {year} {2014})}\BibitemShut {NoStop}%
\bibitem [{\citenamefont {Rudner}\ and\ \citenamefont {Lindner}(2020)}]{Rudner2020}%
  \BibitemOpen
  \bibfield  {author} {\bibinfo {author} {\bibfnamefont {M.~S.}\ \bibnamefont {Rudner}}\ and\ \bibinfo {author} {\bibfnamefont {N.~H.}\ \bibnamefont {Lindner}},\ }\bibfield  {title} {\bibinfo {title} {Band structure engineering and non-equilibrium dynamics in {F}loquet topological insulators},\ }\href {https://doi.org/10.1038/s42254-020-0170-z} {\bibfield  {journal} {\bibinfo  {journal} {Nat. Rev. Phys.}\ }\textbf {\bibinfo {volume} {2}},\ \bibinfo {pages} {229} (\bibinfo {year} {2020})}\BibitemShut {NoStop}%
\bibitem [{\citenamefont {Castro}\ \emph {et~al.}(2022)\citenamefont {Castro}, \citenamefont {De~Giovannini}, \citenamefont {Sato}, \citenamefont {H\"ubener},\ and\ \citenamefont {Rubio}}]{PhysRevResearch.4.033213}%
  \BibitemOpen
  \bibfield  {author} {\bibinfo {author} {\bibfnamefont {A.}~\bibnamefont {Castro}}, \bibinfo {author} {\bibfnamefont {U.}~\bibnamefont {De~Giovannini}}, \bibinfo {author} {\bibfnamefont {S.~A.}\ \bibnamefont {Sato}}, \bibinfo {author} {\bibfnamefont {H.}~\bibnamefont {H\"ubener}},\ and\ \bibinfo {author} {\bibfnamefont {A.}~\bibnamefont {Rubio}},\ }\bibfield  {title} {\bibinfo {title} {Floquet engineering the band structure of materials with optimal control theory},\ }\href {https://doi.org/10.1103/PhysRevResearch.4.033213} {\bibfield  {journal} {\bibinfo  {journal} {Phys. Rev. Res.}\ }\textbf {\bibinfo {volume} {4}},\ \bibinfo {pages} {033213} (\bibinfo {year} {2022})}\BibitemShut {NoStop}%
\bibitem [{\citenamefont {Li}\ \emph {et~al.}(2020)\citenamefont {Li}, \citenamefont {Fertig},\ and\ \citenamefont {Seradjeh}}]{PhysRevResearch.2.043275}%
  \BibitemOpen
  \bibfield  {author} {\bibinfo {author} {\bibfnamefont {Y.}~\bibnamefont {Li}}, \bibinfo {author} {\bibfnamefont {H.~A.}\ \bibnamefont {Fertig}},\ and\ \bibinfo {author} {\bibfnamefont {B.}~\bibnamefont {Seradjeh}},\ }\bibfield  {title} {\bibinfo {title} {Floquet-engineered topological flat bands in irradiated twisted bilayer graphene},\ }\href {https://doi.org/10.1103/PhysRevResearch.2.043275} {\bibfield  {journal} {\bibinfo  {journal} {Phys. Rev. Res.}\ }\textbf {\bibinfo {volume} {2}},\ \bibinfo {pages} {043275} (\bibinfo {year} {2020})}\BibitemShut {NoStop}%
\bibitem [{\citenamefont {Wang}\ \emph {et~al.}(2013)\citenamefont {Wang}, \citenamefont {Steinberg}, \citenamefont {Jarillo-Herrero},\ and\ \citenamefont {Gedik}}]{doi:10.1126/science.1239834}%
  \BibitemOpen
  \bibfield  {author} {\bibinfo {author} {\bibfnamefont {Y.~H.}\ \bibnamefont {Wang}}, \bibinfo {author} {\bibfnamefont {H.}~\bibnamefont {Steinberg}}, \bibinfo {author} {\bibfnamefont {P.}~\bibnamefont {Jarillo-Herrero}},\ and\ \bibinfo {author} {\bibfnamefont {N.}~\bibnamefont {Gedik}},\ }\bibfield  {title} {\bibinfo {title} {Observation of {F}loquet-{B}loch states on the surface of a topological insulator},\ }\href {https://doi.org/10.1126/science.1239834} {\bibfield  {journal} {\bibinfo  {journal} {Science}\ }\textbf {\bibinfo {volume} {342}},\ \bibinfo {pages} {453} (\bibinfo {year} {2013})}\BibitemShut {NoStop}%
\bibitem [{\citenamefont {Boschini}\ \emph {et~al.}(2024)\citenamefont {Boschini}, \citenamefont {Zonno},\ and\ \citenamefont {Damascelli}}]{RevModPhys.96.015003}%
  \BibitemOpen
  \bibfield  {author} {\bibinfo {author} {\bibfnamefont {F.}~\bibnamefont {Boschini}}, \bibinfo {author} {\bibfnamefont {M.}~\bibnamefont {Zonno}},\ and\ \bibinfo {author} {\bibfnamefont {A.}~\bibnamefont {Damascelli}},\ }\bibfield  {title} {\bibinfo {title} {Time-resolved {ARPES} studies of quantum materials},\ }\href {https://doi.org/10.1103/RevModPhys.96.015003} {\bibfield  {journal} {\bibinfo  {journal} {Rev. Mod. Phys.}\ }\textbf {\bibinfo {volume} {96}},\ \bibinfo {pages} {015003} (\bibinfo {year} {2024})}\BibitemShut {NoStop}%
\bibitem [{\citenamefont {Lindner}\ \emph {et~al.}(2011)\citenamefont {Lindner}, \citenamefont {Refael},\ and\ \citenamefont {Galitski}}]{Lindner2011}%
  \BibitemOpen
  \bibfield  {author} {\bibinfo {author} {\bibfnamefont {N.~H.}\ \bibnamefont {Lindner}}, \bibinfo {author} {\bibfnamefont {G.}~\bibnamefont {Refael}},\ and\ \bibinfo {author} {\bibfnamefont {V.}~\bibnamefont {Galitski}},\ }\bibfield  {title} {\bibinfo {title} {Floquet topological insulator in semiconductor quantum wells},\ }\href {https://doi.org/10.1038/nphys1926} {\bibfield  {journal} {\bibinfo  {journal} {Nature Physics}\ }\textbf {\bibinfo {volume} {7}},\ \bibinfo {pages} {490} (\bibinfo {year} {2011})}\BibitemShut {NoStop}%
\bibitem [{\citenamefont {Dehghani}\ and\ \citenamefont {Mitra}(2015)}]{PhysRevB.92.165111}%
  \BibitemOpen
  \bibfield  {author} {\bibinfo {author} {\bibfnamefont {H.}~\bibnamefont {Dehghani}}\ and\ \bibinfo {author} {\bibfnamefont {A.}~\bibnamefont {Mitra}},\ }\bibfield  {title} {\bibinfo {title} {Optical {H}all conductivity of a {F}loquet topological insulator},\ }\href {https://doi.org/10.1103/PhysRevB.92.165111} {\bibfield  {journal} {\bibinfo  {journal} {Phys. Rev. B}\ }\textbf {\bibinfo {volume} {92}},\ \bibinfo {pages} {165111} (\bibinfo {year} {2015})}\BibitemShut {NoStop}%
\bibitem [{\citenamefont {H{\"u}bener}\ \emph {et~al.}(2017)\citenamefont {H{\"u}bener}, \citenamefont {Sentef}, \citenamefont {De~Giovannini}, \citenamefont {Kemper},\ and\ \citenamefont {Rubio}}]{Hubener2017}%
  \BibitemOpen
  \bibfield  {author} {\bibinfo {author} {\bibfnamefont {H.}~\bibnamefont {H{\"u}bener}}, \bibinfo {author} {\bibfnamefont {M.~A.}\ \bibnamefont {Sentef}}, \bibinfo {author} {\bibfnamefont {U.}~\bibnamefont {De~Giovannini}}, \bibinfo {author} {\bibfnamefont {A.~F.}\ \bibnamefont {Kemper}},\ and\ \bibinfo {author} {\bibfnamefont {A.}~\bibnamefont {Rubio}},\ }\bibfield  {title} {\bibinfo {title} {Creating stable {F}loquet-{W}eyl semimetals by laser-driving of 3d {D}irac materials},\ }\href {https://doi.org/10.1038/ncomms13940} {\bibfield  {journal} {\bibinfo  {journal} {Nature Communications}\ }\textbf {\bibinfo {volume} {8}},\ \bibinfo {pages} {13940} (\bibinfo {year} {2017})}\BibitemShut {NoStop}%
\bibitem [{\citenamefont {Kitamura}\ and\ \citenamefont {Aoki}(2022)}]{Kitamura2022}%
  \BibitemOpen
  \bibfield  {author} {\bibinfo {author} {\bibfnamefont {S.}~\bibnamefont {Kitamura}}\ and\ \bibinfo {author} {\bibfnamefont {H.}~\bibnamefont {Aoki}},\ }\bibfield  {title} {\bibinfo {title} {Floquet topological superconductivity induced by chiral many-body interaction},\ }\href {https://doi.org/10.1038/s42005-022-00936-w} {\bibfield  {journal} {\bibinfo  {journal} {Communications Physics}\ }\textbf {\bibinfo {volume} {5}},\ \bibinfo {pages} {174} (\bibinfo {year} {2022})}\BibitemShut {NoStop}%
\bibitem [{\citenamefont {Esin}\ \emph {et~al.}(2020)\citenamefont {Esin}, \citenamefont {Rudner},\ and\ \citenamefont {Lindner}}]{doi:10.1126/sciadv.aay4922}%
  \BibitemOpen
  \bibfield  {author} {\bibinfo {author} {\bibfnamefont {I.}~\bibnamefont {Esin}}, \bibinfo {author} {\bibfnamefont {M.~S.}\ \bibnamefont {Rudner}},\ and\ \bibinfo {author} {\bibfnamefont {N.~H.}\ \bibnamefont {Lindner}},\ }\bibfield  {title} {\bibinfo {title} {Floquet metal-to-insulator phase transitions in semiconductor nanowires},\ }\href {https://doi.org/10.1126/sciadv.aay4922} {\bibfield  {journal} {\bibinfo  {journal} {Science Advances}\ }\textbf {\bibinfo {volume} {6}},\ \bibinfo {pages} {eaay4922} (\bibinfo {year} {2020})}\BibitemShut {NoStop}%
\bibitem [{\citenamefont {Esin}\ \emph {et~al.}(2021)\citenamefont {Esin}, \citenamefont {Gupta}, \citenamefont {Berg}, \citenamefont {Rudner},\ and\ \citenamefont {Lindner}}]{Esin2021}%
  \BibitemOpen
  \bibfield  {author} {\bibinfo {author} {\bibfnamefont {I.}~\bibnamefont {Esin}}, \bibinfo {author} {\bibfnamefont {G.~K.}\ \bibnamefont {Gupta}}, \bibinfo {author} {\bibfnamefont {E.}~\bibnamefont {Berg}}, \bibinfo {author} {\bibfnamefont {M.~S.}\ \bibnamefont {Rudner}},\ and\ \bibinfo {author} {\bibfnamefont {N.~H.}\ \bibnamefont {Lindner}},\ }\bibfield  {title} {\bibinfo {title} {Electronic {F}loquet gyro-liquid crystal},\ }\href {https://doi.org/10.1038/s41467-021-25511-9} {\bibfield  {journal} {\bibinfo  {journal} {Nature Communications}\ }\textbf {\bibinfo {volume} {12}},\ \bibinfo {pages} {5299} (\bibinfo {year} {2021})}\BibitemShut {NoStop}%
\bibitem [{\citenamefont {Fausti}\ \emph {et~al.}(2011)\citenamefont {Fausti}, \citenamefont {Tobey}, \citenamefont {Dean}, \citenamefont {Kaiser}, \citenamefont {Dienst}, \citenamefont {Hoffmann}, \citenamefont {Pyon}, \citenamefont {Takayama}, \citenamefont {Takagi},\ and\ \citenamefont {Cavalleri}}]{Fausti2011}%
  \BibitemOpen
  \bibfield  {author} {\bibinfo {author} {\bibfnamefont {D.}~\bibnamefont {Fausti}}, \bibinfo {author} {\bibfnamefont {R.~I.}\ \bibnamefont {Tobey}}, \bibinfo {author} {\bibfnamefont {N.}~\bibnamefont {Dean}}, \bibinfo {author} {\bibfnamefont {S.}~\bibnamefont {Kaiser}}, \bibinfo {author} {\bibfnamefont {A.}~\bibnamefont {Dienst}}, \bibinfo {author} {\bibfnamefont {M.~C.}\ \bibnamefont {Hoffmann}}, \bibinfo {author} {\bibfnamefont {S.}~\bibnamefont {Pyon}}, \bibinfo {author} {\bibfnamefont {T.}~\bibnamefont {Takayama}}, \bibinfo {author} {\bibfnamefont {H.}~\bibnamefont {Takagi}},\ and\ \bibinfo {author} {\bibfnamefont {A.}~\bibnamefont {Cavalleri}},\ }\bibfield  {title} {\bibinfo {title} {Light-induced superconductivity in a stripe-ordered cuprate},\ }\href {https://doi.org/10.1126/science.1197294} {\bibfield  {journal} {\bibinfo  {journal} {Science}\ }\textbf {\bibinfo {volume} {331}},\ \bibinfo {pages} {189} (\bibinfo {year} {2011})}\BibitemShut {NoStop}%
\bibitem [{\citenamefont {Biswas}\ \emph {et~al.}(2020)\citenamefont {Biswas}, \citenamefont {Mishra}, \citenamefont {Rao},\ and\ \citenamefont {Kundu}}]{PhysRevB.102.155428}%
  \BibitemOpen
  \bibfield  {author} {\bibinfo {author} {\bibfnamefont {S.}~\bibnamefont {Biswas}}, \bibinfo {author} {\bibfnamefont {T.}~\bibnamefont {Mishra}}, \bibinfo {author} {\bibfnamefont {S.}~\bibnamefont {Rao}},\ and\ \bibinfo {author} {\bibfnamefont {A.}~\bibnamefont {Kundu}},\ }\bibfield  {title} {\bibinfo {title} {Chiral {L}uttinger liquids in graphene tuned by irradiation},\ }\href {https://doi.org/10.1103/PhysRevB.102.155428} {\bibfield  {journal} {\bibinfo  {journal} {Phys. Rev. B}\ }\textbf {\bibinfo {volume} {102}},\ \bibinfo {pages} {155428} (\bibinfo {year} {2020})}\BibitemShut {NoStop}%
\bibitem [{\citenamefont {Fazzini}\ \emph {et~al.}(2021)\citenamefont {Fazzini}, \citenamefont {Chudzinski}, \citenamefont {Dauer}, \citenamefont {Schneider},\ and\ \citenamefont {Eggert}}]{PhysRevLett.126.243401}%
  \BibitemOpen
  \bibfield  {author} {\bibinfo {author} {\bibfnamefont {S.}~\bibnamefont {Fazzini}}, \bibinfo {author} {\bibfnamefont {P.}~\bibnamefont {Chudzinski}}, \bibinfo {author} {\bibfnamefont {C.}~\bibnamefont {Dauer}}, \bibinfo {author} {\bibfnamefont {I.}~\bibnamefont {Schneider}},\ and\ \bibinfo {author} {\bibfnamefont {S.}~\bibnamefont {Eggert}},\ }\bibfield  {title} {\bibinfo {title} {Nonequilibrium {F}loquet steady states of time-periodic driven {L}uttinger liquids},\ }\href {https://doi.org/10.1103/PhysRevLett.126.243401} {\bibfield  {journal} {\bibinfo  {journal} {Phys. Rev. Lett.}\ }\textbf {\bibinfo {volume} {126}},\ \bibinfo {pages} {243401} (\bibinfo {year} {2021})}\BibitemShut {NoStop}%
\bibitem [{\citenamefont {D'Alessio}\ and\ \citenamefont {Rigol}(2015)}]{DAlessio2015}%
  \BibitemOpen
  \bibfield  {author} {\bibinfo {author} {\bibfnamefont {L.}~\bibnamefont {D'Alessio}}\ and\ \bibinfo {author} {\bibfnamefont {M.}~\bibnamefont {Rigol}},\ }\bibfield  {title} {\bibinfo {title} {Dynamical preparation of {F}loquet {C}hern insulators},\ }\href {https://doi.org/10.1038/ncomms9336} {\bibfield  {journal} {\bibinfo  {journal} {Nature Communications}\ }\textbf {\bibinfo {volume} {6}},\ \bibinfo {pages} {8336} (\bibinfo {year} {2015})}\BibitemShut {NoStop}%
\bibitem [{\citenamefont {Takahashi}\ \emph {et~al.}(2025)\citenamefont {Takahashi}, \citenamefont {Miyamoto}, \citenamefont {Kuroki},\ and\ \citenamefont {Kaneko}}]{PhysRevB.111.125104}%
  \BibitemOpen
  \bibfield  {author} {\bibinfo {author} {\bibfnamefont {Y.}~\bibnamefont {Takahashi}}, \bibinfo {author} {\bibfnamefont {H.}~\bibnamefont {Miyamoto}}, \bibinfo {author} {\bibfnamefont {K.}~\bibnamefont {Kuroki}},\ and\ \bibinfo {author} {\bibfnamefont {T.}~\bibnamefont {Kaneko}},\ }\bibfield  {title} {\bibinfo {title} {Floquet engineering of effective pairing interactions in a doped band insulator},\ }\href {https://doi.org/10.1103/PhysRevB.111.125104} {\bibfield  {journal} {\bibinfo  {journal} {Phys. Rev. B}\ }\textbf {\bibinfo {volume} {111}},\ \bibinfo {pages} {125104} (\bibinfo {year} {2025})}\BibitemShut {NoStop}%
\bibitem [{\citenamefont {\ifmmode \check{C}\else \v{C}\fi{}ade\ifmmode~\check{z}\else \v{z}\fi{}}\ \emph {et~al.}(2017)\citenamefont {\ifmmode \check{C}\else \v{C}\fi{}ade\ifmmode~\check{z}\else \v{z}\fi{}}, \citenamefont {Mondaini},\ and\ \citenamefont {Sacramento}}]{PhysRevB.96.144301}%
  \BibitemOpen
  \bibfield  {author} {\bibinfo {author} {\bibfnamefont {T.}~\bibnamefont {\ifmmode \check{C}\else \v{C}\fi{}ade\ifmmode~\check{z}\else \v{z}\fi{}}}, \bibinfo {author} {\bibfnamefont {R.}~\bibnamefont {Mondaini}},\ and\ \bibinfo {author} {\bibfnamefont {P.~D.}\ \bibnamefont {Sacramento}},\ }\bibfield  {title} {\bibinfo {title} {Dynamical localization and the effects of aperiodicity in {F}loquet systems},\ }\href {https://doi.org/10.1103/PhysRevB.96.144301} {\bibfield  {journal} {\bibinfo  {journal} {Phys. Rev. B}\ }\textbf {\bibinfo {volume} {96}},\ \bibinfo {pages} {144301} (\bibinfo {year} {2017})}\BibitemShut {NoStop}%
\bibitem [{\citenamefont {Thakurathi}\ \emph {et~al.}(2014)\citenamefont {Thakurathi}, \citenamefont {Sengupta},\ and\ \citenamefont {Sen}}]{thakurathi2014}%
  \BibitemOpen
  \bibfield  {author} {\bibinfo {author} {\bibfnamefont {M.}~\bibnamefont {Thakurathi}}, \bibinfo {author} {\bibfnamefont {K.}~\bibnamefont {Sengupta}},\ and\ \bibinfo {author} {\bibfnamefont {D.}~\bibnamefont {Sen}},\ }\bibfield  {title} {\bibinfo {title} {Majorana edge modes in the {K}itaev model},\ }\href {https://doi.org/10.1103/PhysRevB.89.235434} {\bibfield  {journal} {\bibinfo  {journal} {Phys. Rev. B}\ }\textbf {\bibinfo {volume} {89}},\ \bibinfo {pages} {235434} (\bibinfo {year} {2014})}\BibitemShut {NoStop}%
\bibitem [{\citenamefont {Seshadri}\ and\ \citenamefont {Sen}(2022)}]{seshadri2022}%
  \BibitemOpen
  \bibfield  {author} {\bibinfo {author} {\bibfnamefont {R.}~\bibnamefont {Seshadri}}\ and\ \bibinfo {author} {\bibfnamefont {D.}~\bibnamefont {Sen}},\ }\bibfield  {title} {\bibinfo {title} {Engineering floquet topological phases using elliptically polarized light},\ }\href {https://doi.org/10.1103/PhysRevB.106.245401} {\bibfield  {journal} {\bibinfo  {journal} {Phys. Rev. B}\ }\textbf {\bibinfo {volume} {106}},\ \bibinfo {pages} {245401} (\bibinfo {year} {2022})}\BibitemShut {NoStop}%
\bibitem [{\citenamefont {Martin}\ \emph {et~al.}(2017)\citenamefont {Martin}, \citenamefont {Refael},\ and\ \citenamefont {Halperin}}]{PhysRevX.7.041008}%
  \BibitemOpen
  \bibfield  {author} {\bibinfo {author} {\bibfnamefont {I.}~\bibnamefont {Martin}}, \bibinfo {author} {\bibfnamefont {G.}~\bibnamefont {Refael}},\ and\ \bibinfo {author} {\bibfnamefont {B.}~\bibnamefont {Halperin}},\ }\bibfield  {title} {\bibinfo {title} {Topological frequency conversion in strongly driven quantum systems},\ }\href {https://doi.org/10.1103/PhysRevX.7.041008} {\bibfield  {journal} {\bibinfo  {journal} {Phys. Rev. X}\ }\textbf {\bibinfo {volume} {7}},\ \bibinfo {pages} {041008} (\bibinfo {year} {2017})}\BibitemShut {NoStop}%
\bibitem [{\citenamefont {Vinjamuri}\ \emph {et~al.}(2026)\citenamefont {Vinjamuri}, \citenamefont {Dubey},\ and\ \citenamefont {Das}}]{vinjamuri2026mixedfloquetlatticemodel}%
  \BibitemOpen
  \bibfield  {author} {\bibinfo {author} {\bibfnamefont {G.}~\bibnamefont {Vinjamuri}}, \bibinfo {author} {\bibfnamefont {A.}~\bibnamefont {Dubey}},\ and\ \bibinfo {author} {\bibfnamefont {A.}~\bibnamefont {Das}},\ }\bibfield  {title} {\bibinfo {title} {Mixed {F}loquet lattice model for gapless topology},\ }\href {https://arxiv.org/abs/2606.20378} {\bibfield  {journal} {\bibinfo  {journal} {arXiv:2509.19437}\ } (\bibinfo {year} {2026})}\BibitemShut {NoStop}%
\bibitem [{\citenamefont {Dubey}\ \emph {et~al.}(2023)\citenamefont {Dubey}, \citenamefont {Biswas},\ and\ \citenamefont {Kundu}}]{PhysRevB.108.085433}%
  \BibitemOpen
  \bibfield  {author} {\bibinfo {author} {\bibfnamefont {A.}~\bibnamefont {Dubey}}, \bibinfo {author} {\bibfnamefont {S.}~\bibnamefont {Biswas}},\ and\ \bibinfo {author} {\bibfnamefont {A.}~\bibnamefont {Kundu}},\ }\bibfield  {title} {\bibinfo {title} {Operator correlations in a quenched non-{H}ermitian {L}uttinger liquid},\ }\href {https://doi.org/10.1103/PhysRevB.108.085433} {\bibfield  {journal} {\bibinfo  {journal} {Phys. Rev. B}\ }\textbf {\bibinfo {volume} {108}},\ \bibinfo {pages} {085433} (\bibinfo {year} {2023})}\BibitemShut {NoStop}%
\bibitem [{\citenamefont {Neyenhuis}\ \emph {et~al.}(2017)\citenamefont {Neyenhuis}, \citenamefont {Zhang}, \citenamefont {Hess}, \citenamefont {Smith}, \citenamefont {Lee}, \citenamefont {Richerme}, \citenamefont {Gong}, \citenamefont {Gorshkov},\ and\ \citenamefont {Monroe}}]{doi:10.1126/sciadv.1700672}%
  \BibitemOpen
  \bibfield  {author} {\bibinfo {author} {\bibfnamefont {B.}~\bibnamefont {Neyenhuis}}, \bibinfo {author} {\bibfnamefont {J.}~\bibnamefont {Zhang}}, \bibinfo {author} {\bibfnamefont {P.~W.}\ \bibnamefont {Hess}}, \bibinfo {author} {\bibfnamefont {J.}~\bibnamefont {Smith}}, \bibinfo {author} {\bibfnamefont {A.~C.}\ \bibnamefont {Lee}}, \bibinfo {author} {\bibfnamefont {P.}~\bibnamefont {Richerme}}, \bibinfo {author} {\bibfnamefont {Z.-X.}\ \bibnamefont {Gong}}, \bibinfo {author} {\bibfnamefont {A.~V.}\ \bibnamefont {Gorshkov}},\ and\ \bibinfo {author} {\bibfnamefont {C.}~\bibnamefont {Monroe}},\ }\bibfield  {title} {\bibinfo {title} {Observation of prethermalization in long-range interacting spin chains},\ }\href {https://doi.org/10.1126/sciadv.1700672} {\bibfield  {journal} {\bibinfo  {journal} {Science Advances}\ }\textbf {\bibinfo {volume} {3}},\ \bibinfo {pages} {e1700672} (\bibinfo {year} {2017})}\BibitemShut {NoStop}%
\bibitem [{\citenamefont {Trotzky}\ \emph {et~al.}(2012)\citenamefont {Trotzky}, \citenamefont {Chen}, \citenamefont {Flesch}, \citenamefont {McCulloch}, \citenamefont {Schollw{\"o}ck}, \citenamefont {Eisert},\ and\ \citenamefont {Bloch}}]{Trotzky2012}%
  \BibitemOpen
  \bibfield  {author} {\bibinfo {author} {\bibfnamefont {S.}~\bibnamefont {Trotzky}}, \bibinfo {author} {\bibfnamefont {Y.-A.}\ \bibnamefont {Chen}}, \bibinfo {author} {\bibfnamefont {A.}~\bibnamefont {Flesch}}, \bibinfo {author} {\bibfnamefont {I.~P.}\ \bibnamefont {McCulloch}}, \bibinfo {author} {\bibfnamefont {U.}~\bibnamefont {Schollw{\"o}ck}}, \bibinfo {author} {\bibfnamefont {J.}~\bibnamefont {Eisert}},\ and\ \bibinfo {author} {\bibfnamefont {I.}~\bibnamefont {Bloch}},\ }\bibfield  {title} {\bibinfo {title} {Probing the relaxation towards equilibrium in an isolated strongly correlated one-dimensional {B}ose gas},\ }\href {https://doi.org/10.1038/nphys2232} {\bibfield  {journal} {\bibinfo  {journal} {Nature Physics}\ }\textbf {\bibinfo {volume} {8}},\ \bibinfo {pages} {325} (\bibinfo {year} {2012})}\BibitemShut {NoStop}%
\bibitem [{\citenamefont {Sun}\ \emph {et~al.}(2018)\citenamefont {Sun}, \citenamefont {Yi}, \citenamefont {Wang}, \citenamefont {Zhang}, \citenamefont {Sanders}, \citenamefont {Xu}, \citenamefont {Wang}, \citenamefont {Schmiedmayer}, \citenamefont {Deng}, \citenamefont {Liu}, \citenamefont {Chen},\ and\ \citenamefont {Pan}}]{PhysRevLett.121.250403}%
  \BibitemOpen
  \bibfield  {author} {\bibinfo {author} {\bibfnamefont {W.}~\bibnamefont {Sun}}, \bibinfo {author} {\bibfnamefont {C.-R.}\ \bibnamefont {Yi}}, \bibinfo {author} {\bibfnamefont {B.-Z.}\ \bibnamefont {Wang}}, \bibinfo {author} {\bibfnamefont {W.-W.}\ \bibnamefont {Zhang}}, \bibinfo {author} {\bibfnamefont {B.~C.}\ \bibnamefont {Sanders}}, \bibinfo {author} {\bibfnamefont {X.-T.}\ \bibnamefont {Xu}}, \bibinfo {author} {\bibfnamefont {Z.-Y.}\ \bibnamefont {Wang}}, \bibinfo {author} {\bibfnamefont {J.}~\bibnamefont {Schmiedmayer}}, \bibinfo {author} {\bibfnamefont {Y.}~\bibnamefont {Deng}}, \bibinfo {author} {\bibfnamefont {X.-J.}\ \bibnamefont {Liu}}, \bibinfo {author} {\bibfnamefont {S.}~\bibnamefont {Chen}},\ and\ \bibinfo {author} {\bibfnamefont {J.-W.}\ \bibnamefont {Pan}},\ }\bibfield  {title} {\bibinfo {title} {Uncover topology by quantum quench dynamics},\ }\href {https://doi.org/10.1103/PhysRevLett.121.250403} {\bibfield  {journal} {\bibinfo  {journal} {Phys. Rev. Lett.}\ }\textbf {\bibinfo {volume}
  {121}},\ \bibinfo {pages} {250403} (\bibinfo {year} {2018})}\BibitemShut {NoStop}%
\bibitem [{\citenamefont {Mizoguchi}\ \emph {et~al.}(2021)\citenamefont {Mizoguchi}, \citenamefont {Kuno},\ and\ \citenamefont {Hatsugai}}]{PhysRevLett.126.016802}%
  \BibitemOpen
  \bibfield  {author} {\bibinfo {author} {\bibfnamefont {T.}~\bibnamefont {Mizoguchi}}, \bibinfo {author} {\bibfnamefont {Y.}~\bibnamefont {Kuno}},\ and\ \bibinfo {author} {\bibfnamefont {Y.}~\bibnamefont {Hatsugai}},\ }\bibfield  {title} {\bibinfo {title} {Detecting bulk topology of quadrupolar phase from quench dynamics},\ }\href {https://doi.org/10.1103/PhysRevLett.126.016802} {\bibfield  {journal} {\bibinfo  {journal} {Phys. Rev. Lett.}\ }\textbf {\bibinfo {volume} {126}},\ \bibinfo {pages} {016802} (\bibinfo {year} {2021})}\BibitemShut {NoStop}%
\bibitem [{\citenamefont {Yang}\ \emph {et~al.}(2018)\citenamefont {Yang}, \citenamefont {Li},\ and\ \citenamefont {Chen}}]{PhysRevB.97.060304}%
  \BibitemOpen
  \bibfield  {author} {\bibinfo {author} {\bibfnamefont {C.}~\bibnamefont {Yang}}, \bibinfo {author} {\bibfnamefont {L.}~\bibnamefont {Li}},\ and\ \bibinfo {author} {\bibfnamefont {S.}~\bibnamefont {Chen}},\ }\bibfield  {title} {\bibinfo {title} {Dynamical topological invariant after a quantum quench},\ }\href {https://doi.org/10.1103/PhysRevB.97.060304} {\bibfield  {journal} {\bibinfo  {journal} {Phys. Rev. B}\ }\textbf {\bibinfo {volume} {97}},\ \bibinfo {pages} {060304(R)} (\bibinfo {year} {2018})}\BibitemShut {NoStop}%
\bibitem [{\citenamefont {Pastori}\ \emph {et~al.}(2020)\citenamefont {Pastori}, \citenamefont {Barbarino},\ and\ \citenamefont {Budich}}]{PhysRevResearch.2.033259}%
  \BibitemOpen
  \bibfield  {author} {\bibinfo {author} {\bibfnamefont {L.}~\bibnamefont {Pastori}}, \bibinfo {author} {\bibfnamefont {S.}~\bibnamefont {Barbarino}},\ and\ \bibinfo {author} {\bibfnamefont {J.~C.}\ \bibnamefont {Budich}},\ }\bibfield  {title} {\bibinfo {title} {Signatures of topology in quantum quench dynamics and their interrelation},\ }\href {https://doi.org/10.1103/PhysRevResearch.2.033259} {\bibfield  {journal} {\bibinfo  {journal} {Phys. Rev. Res.}\ }\textbf {\bibinfo {volume} {2}},\ \bibinfo {pages} {033259} (\bibinfo {year} {2020})}\BibitemShut {NoStop}%
\bibitem [{\citenamefont {Bragan\ifmmode~\mbox{\c{c}}\else \c{c}\fi{}a}\ \emph {et~al.}(2021)\citenamefont {Bragan\ifmmode~\mbox{\c{c}}\else \c{c}\fi{}a}, \citenamefont {Cavalcante}, \citenamefont {Pereira},\ and\ \citenamefont {Aguiar}}]{PhysRevB.103.125152}%
  \BibitemOpen
  \bibfield  {author} {\bibinfo {author} {\bibfnamefont {H.}~\bibnamefont {Bragan\ifmmode~\mbox{\c{c}}\else \c{c}\fi{}a}}, \bibinfo {author} {\bibfnamefont {M.~F.}\ \bibnamefont {Cavalcante}}, \bibinfo {author} {\bibfnamefont {R.~G.}\ \bibnamefont {Pereira}},\ and\ \bibinfo {author} {\bibfnamefont {M.~C.~O.}\ \bibnamefont {Aguiar}},\ }\bibfield  {title} {\bibinfo {title} {Quench dynamics and relaxation of a spin coupled to interacting leads},\ }\href {https://doi.org/10.1103/PhysRevB.103.125152} {\bibfield  {journal} {\bibinfo  {journal} {Phys. Rev. B}\ }\textbf {\bibinfo {volume} {103}},\ \bibinfo {pages} {125152} (\bibinfo {year} {2021})}\BibitemShut {NoStop}%
\bibitem [{\citenamefont {Eckstein}\ \emph {et~al.}(2009)\citenamefont {Eckstein}, \citenamefont {Kollar},\ and\ \citenamefont {Werner}}]{PhysRevLett.103.056403}%
  \BibitemOpen
  \bibfield  {author} {\bibinfo {author} {\bibfnamefont {M.}~\bibnamefont {Eckstein}}, \bibinfo {author} {\bibfnamefont {M.}~\bibnamefont {Kollar}},\ and\ \bibinfo {author} {\bibfnamefont {P.}~\bibnamefont {Werner}},\ }\bibfield  {title} {\bibinfo {title} {Thermalization after an interaction quench in the hubbard model},\ }\href {https://doi.org/10.1103/PhysRevLett.103.056403} {\bibfield  {journal} {\bibinfo  {journal} {Phys. Rev. Lett.}\ }\textbf {\bibinfo {volume} {103}},\ \bibinfo {pages} {056403} (\bibinfo {year} {2009})}\BibitemShut {NoStop}%
\bibitem [{\citenamefont {Sharma}\ \emph {et~al.}(2015)\citenamefont {Sharma}, \citenamefont {Suzuki},\ and\ \citenamefont {Dutta}}]{PhysRevB.92.104306}%
  \BibitemOpen
  \bibfield  {author} {\bibinfo {author} {\bibfnamefont {S.}~\bibnamefont {Sharma}}, \bibinfo {author} {\bibfnamefont {S.}~\bibnamefont {Suzuki}},\ and\ \bibinfo {author} {\bibfnamefont {A.}~\bibnamefont {Dutta}},\ }\bibfield  {title} {\bibinfo {title} {Quenches and dynamical phase transitions in a nonintegrable quantum {I}sing model},\ }\href {https://doi.org/10.1103/PhysRevB.92.104306} {\bibfield  {journal} {\bibinfo  {journal} {Phys. Rev. B}\ }\textbf {\bibinfo {volume} {92}},\ \bibinfo {pages} {104306} (\bibinfo {year} {2015})}\BibitemShut {NoStop}%
\bibitem [{\citenamefont {Sharma}\ \emph {et~al.}(2016)\citenamefont {Sharma}, \citenamefont {Divakaran}, \citenamefont {Polkovnikov},\ and\ \citenamefont {Dutta}}]{PhysRevB.93.144306}%
  \BibitemOpen
  \bibfield  {author} {\bibinfo {author} {\bibfnamefont {S.}~\bibnamefont {Sharma}}, \bibinfo {author} {\bibfnamefont {U.}~\bibnamefont {Divakaran}}, \bibinfo {author} {\bibfnamefont {A.}~\bibnamefont {Polkovnikov}},\ and\ \bibinfo {author} {\bibfnamefont {A.}~\bibnamefont {Dutta}},\ }\bibfield  {title} {\bibinfo {title} {Slow quenches in a quantum ising chain: Dynamical phase transitions and topology},\ }\href {https://doi.org/10.1103/PhysRevB.93.144306} {\bibfield  {journal} {\bibinfo  {journal} {Phys. Rev. B}\ }\textbf {\bibinfo {volume} {93}},\ \bibinfo {pages} {144306} (\bibinfo {year} {2016})}\BibitemShut {NoStop}%
\bibitem [{\citenamefont {Ding}(2020)}]{PhysRevB.102.060409}%
  \BibitemOpen
  \bibfield  {author} {\bibinfo {author} {\bibfnamefont {C.}~\bibnamefont {Ding}},\ }\bibfield  {title} {\bibinfo {title} {Dynamical quantum phase transition from a critical quantum quench},\ }\href {https://doi.org/10.1103/PhysRevB.102.060409} {\bibfield  {journal} {\bibinfo  {journal} {Phys. Rev. B}\ }\textbf {\bibinfo {volume} {102}},\ \bibinfo {pages} {060409(R)} (\bibinfo {year} {2020})}\BibitemShut {NoStop}%
\bibitem [{\citenamefont {Schachenmayer}\ \emph {et~al.}(2013)\citenamefont {Schachenmayer}, \citenamefont {Lanyon}, \citenamefont {Roos},\ and\ \citenamefont {Daley}}]{PhysRevX.3.031015}%
  \BibitemOpen
  \bibfield  {author} {\bibinfo {author} {\bibfnamefont {J.}~\bibnamefont {Schachenmayer}}, \bibinfo {author} {\bibfnamefont {B.~P.}\ \bibnamefont {Lanyon}}, \bibinfo {author} {\bibfnamefont {C.~F.}\ \bibnamefont {Roos}},\ and\ \bibinfo {author} {\bibfnamefont {A.~J.}\ \bibnamefont {Daley}},\ }\bibfield  {title} {\bibinfo {title} {Entanglement growth in quench dynamics with variable range interactions},\ }\href {https://doi.org/10.1103/PhysRevX.3.031015} {\bibfield  {journal} {\bibinfo  {journal} {Phys. Rev. X}\ }\textbf {\bibinfo {volume} {3}},\ \bibinfo {pages} {031015} (\bibinfo {year} {2013})}\BibitemShut {NoStop}%
\bibitem [{\citenamefont {B\'acsi}\ and\ \citenamefont {D\'ora}(2021)}]{PhysRevB.103.085137}%
  \BibitemOpen
  \bibfield  {author} {\bibinfo {author} {\bibfnamefont {A.}~\bibnamefont {B\'acsi}}\ and\ \bibinfo {author} {\bibfnamefont {B.}~\bibnamefont {D\'ora}},\ }\bibfield  {title} {\bibinfo {title} {Dynamics of entanglement after exceptional quantum quench},\ }\href {https://doi.org/10.1103/PhysRevB.103.085137} {\bibfield  {journal} {\bibinfo  {journal} {Phys. Rev. B}\ }\textbf {\bibinfo {volume} {103}},\ \bibinfo {pages} {085137} (\bibinfo {year} {2021})}\BibitemShut {NoStop}%
\bibitem [{\citenamefont {Rakovszky}\ \emph {et~al.}(2019)\citenamefont {Rakovszky}, \citenamefont {von Keyserlingk},\ and\ \citenamefont {Pollmann}}]{PhysRevB.100.125139}%
  \BibitemOpen
  \bibfield  {author} {\bibinfo {author} {\bibfnamefont {T.}~\bibnamefont {Rakovszky}}, \bibinfo {author} {\bibfnamefont {C.~W.}\ \bibnamefont {von Keyserlingk}},\ and\ \bibinfo {author} {\bibfnamefont {F.}~\bibnamefont {Pollmann}},\ }\bibfield  {title} {\bibinfo {title} {Entanglement growth after inhomogenous quenches},\ }\href {https://doi.org/10.1103/PhysRevB.100.125139} {\bibfield  {journal} {\bibinfo  {journal} {Phys. Rev. B}\ }\textbf {\bibinfo {volume} {100}},\ \bibinfo {pages} {125139} (\bibinfo {year} {2019})}\BibitemShut {NoStop}%
\bibitem [{\citenamefont {Mukherjee}\ \emph {et~al.}(2007)\citenamefont {Mukherjee}, \citenamefont {Divakaran}, \citenamefont {Dutta},\ and\ \citenamefont {Sen}}]{PhysRevB.76.174303}%
  \BibitemOpen
  \bibfield  {author} {\bibinfo {author} {\bibfnamefont {V.}~\bibnamefont {Mukherjee}}, \bibinfo {author} {\bibfnamefont {U.}~\bibnamefont {Divakaran}}, \bibinfo {author} {\bibfnamefont {A.}~\bibnamefont {Dutta}},\ and\ \bibinfo {author} {\bibfnamefont {D.}~\bibnamefont {Sen}},\ }\bibfield  {title} {\bibinfo {title} {Quenching dynamics of a quantum $xy$ spin-$\frac{1}{2}$ chain in a transverse field},\ }\href {https://doi.org/10.1103/PhysRevB.76.174303} {\bibfield  {journal} {\bibinfo  {journal} {Phys. Rev. B}\ }\textbf {\bibinfo {volume} {76}},\ \bibinfo {pages} {174303} (\bibinfo {year} {2007})}\BibitemShut {NoStop}%
\bibitem [{\citenamefont {B{\'a}csi}\ and\ \citenamefont {D{\'o}ra}(2023)}]{Bacsi2023}%
  \BibitemOpen
  \bibfield  {author} {\bibinfo {author} {\bibfnamefont {{\'A}.}~\bibnamefont {B{\'a}csi}}\ and\ \bibinfo {author} {\bibfnamefont {B.}~\bibnamefont {D{\'o}ra}},\ }\bibfield  {title} {\bibinfo {title} {Kibble--{Z}urek scaling due to environment temperature quench in the transverse field ising model},\ }\href {https://doi.org/10.1038/s41598-023-30840-4} {\bibfield  {journal} {\bibinfo  {journal} {Scientific Reports}\ }\textbf {\bibinfo {volume} {13}},\ \bibinfo {pages} {4034} (\bibinfo {year} {2023})}\BibitemShut {NoStop}%
\bibitem [{\citenamefont {Mardanya}\ \emph {et~al.}(2018)\citenamefont {Mardanya}, \citenamefont {Bhattacharya}, \citenamefont {Agarwal},\ and\ \citenamefont {Dutta}}]{PhysRevB.97.115443}%
  \BibitemOpen
  \bibfield  {author} {\bibinfo {author} {\bibfnamefont {S.}~\bibnamefont {Mardanya}}, \bibinfo {author} {\bibfnamefont {U.}~\bibnamefont {Bhattacharya}}, \bibinfo {author} {\bibfnamefont {A.}~\bibnamefont {Agarwal}},\ and\ \bibinfo {author} {\bibfnamefont {A.}~\bibnamefont {Dutta}},\ }\bibfield  {title} {\bibinfo {title} {Dynamics of edge currents in a linearly quenched {H}aldane model},\ }\href {https://doi.org/10.1103/PhysRevB.97.115443} {\bibfield  {journal} {\bibinfo  {journal} {Phys. Rev. B}\ }\textbf {\bibinfo {volume} {97}},\ \bibinfo {pages} {115443} (\bibinfo {year} {2018})}\BibitemShut {NoStop}%
\bibitem [{\citenamefont {Caio}\ \emph {et~al.}(2015)\citenamefont {Caio}, \citenamefont {Cooper},\ and\ \citenamefont {Bhaseen}}]{PhysRevLett.115.236403}%
  \BibitemOpen
  \bibfield  {author} {\bibinfo {author} {\bibfnamefont {M.~D.}\ \bibnamefont {Caio}}, \bibinfo {author} {\bibfnamefont {N.~R.}\ \bibnamefont {Cooper}},\ and\ \bibinfo {author} {\bibfnamefont {M.~J.}\ \bibnamefont {Bhaseen}},\ }\bibfield  {title} {\bibinfo {title} {Quantum quenches in {C}hern insulators},\ }\href {https://doi.org/10.1103/PhysRevLett.115.236403} {\bibfield  {journal} {\bibinfo  {journal} {Phys. Rev. Lett.}\ }\textbf {\bibinfo {volume} {115}},\ \bibinfo {pages} {236403} (\bibinfo {year} {2015})}\BibitemShut {NoStop}%
\bibitem [{\citenamefont {Deng}\ and\ \citenamefont {Yang}(2026)}]{3wcr-sxtz}%
  \BibitemOpen
  \bibfield  {author} {\bibinfo {author} {\bibfnamefont {T.-S.}\ \bibnamefont {Deng}}\ and\ \bibinfo {author} {\bibfnamefont {F.}~\bibnamefont {Yang}},\ }\bibfield  {title} {\bibinfo {title} {Dissipation-driven topological phase transitions in open quantum systems independent of system {H}amiltonian},\ }\href {https://doi.org/10.1103/3wcr-sxtz} {\bibfield  {journal} {\bibinfo  {journal} {Phys. Rev. B}\ }\textbf {\bibinfo {volume} {113}},\ \bibinfo {pages} {024312} (\bibinfo {year} {2026})}\BibitemShut {NoStop}%
\bibitem [{\citenamefont {Dehghani}\ and\ \citenamefont {Mitra}(2016)}]{PhysRevB.93.205437}%
  \BibitemOpen
  \bibfield  {author} {\bibinfo {author} {\bibfnamefont {H.}~\bibnamefont {Dehghani}}\ and\ \bibinfo {author} {\bibfnamefont {A.}~\bibnamefont {Mitra}},\ }\bibfield  {title} {\bibinfo {title} {Occupation probabilities and current densities of bulk and edge states of a {F}loquet topological insulator},\ }\href {https://doi.org/10.1103/PhysRevB.93.205437} {\bibfield  {journal} {\bibinfo  {journal} {Phys. Rev. B}\ }\textbf {\bibinfo {volume} {93}},\ \bibinfo {pages} {205437} (\bibinfo {year} {2016})}\BibitemShut {NoStop}%
\bibitem [{\citenamefont {Matsyshyn}\ \emph {et~al.}(2023)\citenamefont {Matsyshyn}, \citenamefont {Song}, \citenamefont {Villadiego},\ and\ \citenamefont {Shi}}]{PhysRevB.107.195135}%
  \BibitemOpen
  \bibfield  {author} {\bibinfo {author} {\bibfnamefont {O.}~\bibnamefont {Matsyshyn}}, \bibinfo {author} {\bibfnamefont {J.~C.~W.}\ \bibnamefont {Song}}, \bibinfo {author} {\bibfnamefont {I.~S.}\ \bibnamefont {Villadiego}},\ and\ \bibinfo {author} {\bibfnamefont {L.-k.}\ \bibnamefont {Shi}},\ }\bibfield  {title} {\bibinfo {title} {Fermi-{D}irac staircase occupation of {F}loquet bands and current rectification inside the optical gap of metals: An exact approach},\ }\href {https://doi.org/10.1103/PhysRevB.107.195135} {\bibfield  {journal} {\bibinfo  {journal} {Phys. Rev. B}\ }\textbf {\bibinfo {volume} {107}},\ \bibinfo {pages} {195135} (\bibinfo {year} {2023})}\BibitemShut {NoStop}%
\bibitem [{\citenamefont {Seetharam}\ \emph {et~al.}(2015)\citenamefont {Seetharam}, \citenamefont {Bardyn}, \citenamefont {Lindner}, \citenamefont {Rudner},\ and\ \citenamefont {Refael}}]{PhysRevX.5.041050}%
  \BibitemOpen
  \bibfield  {author} {\bibinfo {author} {\bibfnamefont {K.~I.}\ \bibnamefont {Seetharam}}, \bibinfo {author} {\bibfnamefont {C.-E.}\ \bibnamefont {Bardyn}}, \bibinfo {author} {\bibfnamefont {N.~H.}\ \bibnamefont {Lindner}}, \bibinfo {author} {\bibfnamefont {M.~S.}\ \bibnamefont {Rudner}},\ and\ \bibinfo {author} {\bibfnamefont {G.}~\bibnamefont {Refael}},\ }\bibfield  {title} {\bibinfo {title} {Controlled population of {F}loquet-{B}loch states via coupling to {B}ose and {F}ermi baths},\ }\href {https://doi.org/10.1103/PhysRevX.5.041050} {\bibfield  {journal} {\bibinfo  {journal} {Phys. Rev. X}\ }\textbf {\bibinfo {volume} {5}},\ \bibinfo {pages} {041050} (\bibinfo {year} {2015})}\BibitemShut {NoStop}%
\bibitem [{\citenamefont {Kohler}\ \emph {et~al.}(2005)\citenamefont {Kohler}, \citenamefont {Lehmann},\ and\ \citenamefont {Hänggi}}]{KOHLER2005379}%
  \BibitemOpen
  \bibfield  {author} {\bibinfo {author} {\bibfnamefont {S.}~\bibnamefont {Kohler}}, \bibinfo {author} {\bibfnamefont {J.}~\bibnamefont {Lehmann}},\ and\ \bibinfo {author} {\bibfnamefont {P.}~\bibnamefont {Hänggi}},\ }\bibfield  {title} {\bibinfo {title} {Driven quantum transport on the nanoscale},\ }\href {https://doi.org/https://doi.org/10.1016/j.physrep.2004.11.002} {\bibfield  {journal} {\bibinfo  {journal} {Physics Reports}\ }\textbf {\bibinfo {volume} {406}},\ \bibinfo {pages} {379} (\bibinfo {year} {2005})}\BibitemShut {NoStop}%
\bibitem [{\citenamefont {Iadecola}\ and\ \citenamefont {Chamon}(2015)}]{PhysRevB.91.184301}%
  \BibitemOpen
  \bibfield  {author} {\bibinfo {author} {\bibfnamefont {T.}~\bibnamefont {Iadecola}}\ and\ \bibinfo {author} {\bibfnamefont {C.}~\bibnamefont {Chamon}},\ }\bibfield  {title} {\bibinfo {title} {Floquet systems coupled to particle reservoirs},\ }\href {https://doi.org/10.1103/PhysRevB.91.184301} {\bibfield  {journal} {\bibinfo  {journal} {Phys. Rev. B}\ }\textbf {\bibinfo {volume} {91}},\ \bibinfo {pages} {184301} (\bibinfo {year} {2015})}\BibitemShut {NoStop}%
\bibitem [{\citenamefont {Dehghani}\ \emph {et~al.}(2014)\citenamefont {Dehghani}, \citenamefont {Oka},\ and\ \citenamefont {Mitra}}]{PhysRevB.90.195429}%
  \BibitemOpen
  \bibfield  {author} {\bibinfo {author} {\bibfnamefont {H.}~\bibnamefont {Dehghani}}, \bibinfo {author} {\bibfnamefont {T.}~\bibnamefont {Oka}},\ and\ \bibinfo {author} {\bibfnamefont {A.}~\bibnamefont {Mitra}},\ }\bibfield  {title} {\bibinfo {title} {Dissipative {F}loquet topological systems},\ }\href {https://doi.org/10.1103/PhysRevB.90.195429} {\bibfield  {journal} {\bibinfo  {journal} {Phys. Rev. B}\ }\textbf {\bibinfo {volume} {90}},\ \bibinfo {pages} {195429} (\bibinfo {year} {2014})}\BibitemShut {NoStop}%
\bibitem [{\citenamefont {Dehghani}\ \emph {et~al.}(2015)\citenamefont {Dehghani}, \citenamefont {Oka},\ and\ \citenamefont {Mitra}}]{PhysRevB.91.155422}%
  \BibitemOpen
  \bibfield  {author} {\bibinfo {author} {\bibfnamefont {H.}~\bibnamefont {Dehghani}}, \bibinfo {author} {\bibfnamefont {T.}~\bibnamefont {Oka}},\ and\ \bibinfo {author} {\bibfnamefont {A.}~\bibnamefont {Mitra}},\ }\bibfield  {title} {\bibinfo {title} {Out-of-equilibrium electrons and the {H}all conductance of a {F}loquet topological insulator},\ }\href {https://doi.org/10.1103/PhysRevB.91.155422} {\bibfield  {journal} {\bibinfo  {journal} {Phys. Rev. B}\ }\textbf {\bibinfo {volume} {91}},\ \bibinfo {pages} {155422} (\bibinfo {year} {2015})}\BibitemShut {NoStop}%
\bibitem [{\citenamefont {Dubey}\ \emph {et~al.}(2025)\citenamefont {Dubey}, \citenamefont {Kundu},\ and\ \citenamefont {Kundu}}]{PhysRevB.111.035431}%
  \BibitemOpen
  \bibfield  {author} {\bibinfo {author} {\bibfnamefont {A.}~\bibnamefont {Dubey}}, \bibinfo {author} {\bibfnamefont {R.}~\bibnamefont {Kundu}},\ and\ \bibinfo {author} {\bibfnamefont {A.}~\bibnamefont {Kundu}},\ }\bibfield  {title} {\bibinfo {title} {Time-resolved {ARPES} and optical transport properties of irradiated twisted bilayer graphene in a steady state},\ }\href {https://doi.org/10.1103/PhysRevB.111.035431} {\bibfield  {journal} {\bibinfo  {journal} {Phys. Rev. B}\ }\textbf {\bibinfo {volume} {111}},\ \bibinfo {pages} {035431} (\bibinfo {year} {2025})}\BibitemShut {NoStop}%
\bibitem [{\citenamefont {Shi}\ \emph {et~al.}(2024)\citenamefont {Shi}, \citenamefont {Matsyshyn}, \citenamefont {Song},\ and\ \citenamefont {Villadiego}}]{PhysRevLett.132.146402}%
  \BibitemOpen
  \bibfield  {author} {\bibinfo {author} {\bibfnamefont {L.-k.}\ \bibnamefont {Shi}}, \bibinfo {author} {\bibfnamefont {O.}~\bibnamefont {Matsyshyn}}, \bibinfo {author} {\bibfnamefont {J.~C.~W.}\ \bibnamefont {Song}},\ and\ \bibinfo {author} {\bibfnamefont {I.~S.}\ \bibnamefont {Villadiego}},\ }\bibfield  {title} {\bibinfo {title} {Floquet {F}ermi liquid},\ }\href {https://doi.org/10.1103/PhysRevLett.132.146402} {\bibfield  {journal} {\bibinfo  {journal} {Phys. Rev. Lett.}\ }\textbf {\bibinfo {volume} {132}},\ \bibinfo {pages} {146402} (\bibinfo {year} {2024})}\BibitemShut {NoStop}%
\bibitem [{\citenamefont {Kumari}\ \emph {et~al.}(2024)\citenamefont {Kumari}, \citenamefont {Seradjeh},\ and\ \citenamefont {Kundu}}]{PhysRevLett.133.196601}%
  \BibitemOpen
  \bibfield  {author} {\bibinfo {author} {\bibfnamefont {R.}~\bibnamefont {Kumari}}, \bibinfo {author} {\bibfnamefont {B.}~\bibnamefont {Seradjeh}},\ and\ \bibinfo {author} {\bibfnamefont {A.}~\bibnamefont {Kundu}},\ }\bibfield  {title} {\bibinfo {title} {Josephson-current signatures of unpaired {F}loquet {M}ajorana fermions},\ }\href {https://doi.org/10.1103/PhysRevLett.133.196601} {\bibfield  {journal} {\bibinfo  {journal} {Phys. Rev. Lett.}\ }\textbf {\bibinfo {volume} {133}},\ \bibinfo {pages} {196601} (\bibinfo {year} {2024})}\BibitemShut {NoStop}%
\bibitem [{\citenamefont {Rudner}\ \emph {et~al.}(2013)\citenamefont {Rudner}, \citenamefont {Lindner}, \citenamefont {Berg},\ and\ \citenamefont {Levin}}]{PhysRevX.3.031005}%
  \BibitemOpen
  \bibfield  {author} {\bibinfo {author} {\bibfnamefont {M.~S.}\ \bibnamefont {Rudner}}, \bibinfo {author} {\bibfnamefont {N.~H.}\ \bibnamefont {Lindner}}, \bibinfo {author} {\bibfnamefont {E.}~\bibnamefont {Berg}},\ and\ \bibinfo {author} {\bibfnamefont {M.}~\bibnamefont {Levin}},\ }\bibfield  {title} {\bibinfo {title} {Anomalous edge states and the bulk-edge correspondence for periodically driven two-dimensional systems},\ }\href {https://doi.org/10.1103/PhysRevX.3.031005} {\bibfield  {journal} {\bibinfo  {journal} {Phys. Rev. X}\ }\textbf {\bibinfo {volume} {3}},\ \bibinfo {pages} {031005} (\bibinfo {year} {2013})}\BibitemShut {NoStop}%
\bibitem [{\citenamefont {Gritsev}\ and\ \citenamefont {Polkovnikov}(2017)}]{10.21468/SciPostPhys.2.3.021}%
  \BibitemOpen
  \bibfield  {author} {\bibinfo {author} {\bibfnamefont {V.}~\bibnamefont {Gritsev}}\ and\ \bibinfo {author} {\bibfnamefont {A.}~\bibnamefont {Polkovnikov}},\ }\bibfield  {title} {\bibinfo {title} {{Integrable {F}loquet dynamics}},\ }\href {https://doi.org/10.21468/SciPostPhys.2.3.021} {\bibfield  {journal} {\bibinfo  {journal} {SciPost Phys.}\ }\textbf {\bibinfo {volume} {2}},\ \bibinfo {pages} {021} (\bibinfo {year} {2017})}\BibitemShut {NoStop}%
\bibitem [{\citenamefont {Rajak}\ \emph {et~al.}(2019)\citenamefont {Rajak}, \citenamefont {Dana},\ and\ \citenamefont {Dalla~Torre}}]{PhysRevB.100.100302}%
  \BibitemOpen
  \bibfield  {author} {\bibinfo {author} {\bibfnamefont {A.}~\bibnamefont {Rajak}}, \bibinfo {author} {\bibfnamefont {I.}~\bibnamefont {Dana}},\ and\ \bibinfo {author} {\bibfnamefont {E.~G.}\ \bibnamefont {Dalla~Torre}},\ }\bibfield  {title} {\bibinfo {title} {Characterizations of prethermal states in periodically driven many-body systems with unbounded chaotic diffusion},\ }\href {https://doi.org/10.1103/PhysRevB.100.100302} {\bibfield  {journal} {\bibinfo  {journal} {Phys. Rev. B}\ }\textbf {\bibinfo {volume} {100}},\ \bibinfo {pages} {100302(R)} (\bibinfo {year} {2019})}\BibitemShut {NoStop}%
\bibitem [{\citenamefont {Foa~Torres}\ \emph {et~al.}(2014)\citenamefont {Foa~Torres}, \citenamefont {Perez-Piskunow}, \citenamefont {Balseiro},\ and\ \citenamefont {Usaj}}]{PhysRevLett.113.266801}%
  \BibitemOpen
  \bibfield  {author} {\bibinfo {author} {\bibfnamefont {L.~E.~F.}\ \bibnamefont {Foa~Torres}}, \bibinfo {author} {\bibfnamefont {P.~M.}\ \bibnamefont {Perez-Piskunow}}, \bibinfo {author} {\bibfnamefont {C.~A.}\ \bibnamefont {Balseiro}},\ and\ \bibinfo {author} {\bibfnamefont {G.}~\bibnamefont {Usaj}},\ }\bibfield  {title} {\bibinfo {title} {Multiterminal conductance of a {F}loquet topological insulator},\ }\href {https://doi.org/10.1103/PhysRevLett.113.266801} {\bibfield  {journal} {\bibinfo  {journal} {Phys. Rev. Lett.}\ }\textbf {\bibinfo {volume} {113}},\ \bibinfo {pages} {266801} (\bibinfo {year} {2014})}\BibitemShut {NoStop}%
\bibitem [{\citenamefont {Kumari}\ \emph {et~al.}(2026)\citenamefont {Kumari}, \citenamefont {Kulkarni},\ and\ \citenamefont {Dhar}}]{kumari2026quantizedtransportfloquettopological}%
  \BibitemOpen
  \bibfield  {author} {\bibinfo {author} {\bibfnamefont {R.}~\bibnamefont {Kumari}}, \bibinfo {author} {\bibfnamefont {M.}~\bibnamefont {Kulkarni}},\ and\ \bibinfo {author} {\bibfnamefont {A.}~\bibnamefont {Dhar}},\ }\bibfield  {title} {\bibinfo {title} {Quantized transport in {F}loquet topological insulators},\ }\href {https://arxiv.org/abs/2605.13066} {\bibfield  {journal} {\bibinfo  {journal} {arXiv:2605.13066}\ } (\bibinfo {year} {2026})}\BibitemShut {NoStop}%
\bibitem [{\citenamefont {D'Alessio}\ and\ \citenamefont {Rigol}(2014)}]{PhysRevX.4.041048}%
  \BibitemOpen
  \bibfield  {author} {\bibinfo {author} {\bibfnamefont {L.}~\bibnamefont {D'Alessio}}\ and\ \bibinfo {author} {\bibfnamefont {M.}~\bibnamefont {Rigol}},\ }\bibfield  {title} {\bibinfo {title} {Long-time behavior of isolated periodically driven interacting lattice systems},\ }\href {https://doi.org/10.1103/PhysRevX.4.041048} {\bibfield  {journal} {\bibinfo  {journal} {Phys. Rev. X}\ }\textbf {\bibinfo {volume} {4}},\ \bibinfo {pages} {041048} (\bibinfo {year} {2014})}\BibitemShut {NoStop}%
\bibitem [{\citenamefont {Bluhm}\ \emph {et~al.}(2009)\citenamefont {Bluhm}, \citenamefont {Koshnick}, \citenamefont {Bert}, \citenamefont {Huber},\ and\ \citenamefont {Moler}}]{PhysRevLett.102.136802}%
  \BibitemOpen
  \bibfield  {author} {\bibinfo {author} {\bibfnamefont {H.}~\bibnamefont {Bluhm}}, \bibinfo {author} {\bibfnamefont {N.~C.}\ \bibnamefont {Koshnick}}, \bibinfo {author} {\bibfnamefont {J.~A.}\ \bibnamefont {Bert}}, \bibinfo {author} {\bibfnamefont {M.~E.}\ \bibnamefont {Huber}},\ and\ \bibinfo {author} {\bibfnamefont {K.~A.}\ \bibnamefont {Moler}},\ }\bibfield  {title} {\bibinfo {title} {Persistent currents in normal metal rings},\ }\href {https://doi.org/10.1103/PhysRevLett.102.136802} {\bibfield  {journal} {\bibinfo  {journal} {Phys. Rev. Lett.}\ }\textbf {\bibinfo {volume} {102}},\ \bibinfo {pages} {136802} (\bibinfo {year} {2009})}\BibitemShut {NoStop}%
\bibitem [{\citenamefont {Shibata}\ \emph {et~al.}(2015)\citenamefont {Shibata}, \citenamefont {Nomura}, \citenamefont {Kashiwaya}, \citenamefont {Kashiwaya}, \citenamefont {Ishiguro},\ and\ \citenamefont {Takayanagi}}]{shibata2015imaging}%
  \BibitemOpen
  \bibfield  {author} {\bibinfo {author} {\bibfnamefont {Y.}~\bibnamefont {Shibata}}, \bibinfo {author} {\bibfnamefont {S.}~\bibnamefont {Nomura}}, \bibinfo {author} {\bibfnamefont {H.}~\bibnamefont {Kashiwaya}}, \bibinfo {author} {\bibfnamefont {S.}~\bibnamefont {Kashiwaya}}, \bibinfo {author} {\bibfnamefont {R.}~\bibnamefont {Ishiguro}},\ and\ \bibinfo {author} {\bibfnamefont {H.}~\bibnamefont {Takayanagi}},\ }\bibfield  {title} {\bibinfo {title} {Imaging of current density distributions with a {N}b weak-link scanning nano-squid microscope},\ }\href {https://doi.org/10.1038/srep15097} {\bibfield  {journal} {\bibinfo  {journal} {Scientific Reports}\ }\textbf {\bibinfo {volume} {5}},\ \bibinfo {pages} {15097} (\bibinfo {year} {2015})}\BibitemShut {NoStop}%
\bibitem [{\citenamefont {Yap}\ \emph {et~al.}(2017)\citenamefont {Yap}, \citenamefont {Zhou}, \citenamefont {Wang},\ and\ \citenamefont {Gong}}]{PhysRevB.96.165443}%
  \BibitemOpen
  \bibfield  {author} {\bibinfo {author} {\bibfnamefont {H.~H.}\ \bibnamefont {Yap}}, \bibinfo {author} {\bibfnamefont {L.}~\bibnamefont {Zhou}}, \bibinfo {author} {\bibfnamefont {J.-S.}\ \bibnamefont {Wang}},\ and\ \bibinfo {author} {\bibfnamefont {J.}~\bibnamefont {Gong}},\ }\bibfield  {title} {\bibinfo {title} {Computational study of the two-terminal transport of {F}loquet quantum {H}all insulators},\ }\href {https://doi.org/10.1103/PhysRevB.96.165443} {\bibfield  {journal} {\bibinfo  {journal} {Phys. Rev. B}\ }\textbf {\bibinfo {volume} {96}},\ \bibinfo {pages} {165443} (\bibinfo {year} {2017})}\BibitemShut {NoStop}%
\bibitem [{\citenamefont {Kundu}\ \emph {et~al.}(2020)\citenamefont {Kundu}, \citenamefont {Rudner}, \citenamefont {Berg},\ and\ \citenamefont {Lindner}}]{PhysRevB.101.041403}%
  \BibitemOpen
  \bibfield  {author} {\bibinfo {author} {\bibfnamefont {A.}~\bibnamefont {Kundu}}, \bibinfo {author} {\bibfnamefont {M.}~\bibnamefont {Rudner}}, \bibinfo {author} {\bibfnamefont {E.}~\bibnamefont {Berg}},\ and\ \bibinfo {author} {\bibfnamefont {N.~H.}\ \bibnamefont {Lindner}},\ }\bibfield  {title} {\bibinfo {title} {Quantized large-bias current in the anomalous {F}loquet-anderson insulator},\ }\href {https://doi.org/10.1103/PhysRevB.101.041403} {\bibfield  {journal} {\bibinfo  {journal} {Phys. Rev. B}\ }\textbf {\bibinfo {volume} {101}},\ \bibinfo {pages} {041403(R)} (\bibinfo {year} {2020})}\BibitemShut {NoStop}%
\bibitem [{\citenamefont {\"Unal}\ \emph {et~al.}(2019)\citenamefont {\"Unal}, \citenamefont {Eckardt},\ and\ \citenamefont {Slager}}]{PhysRevResearch.1.022003}%
  \BibitemOpen
  \bibfield  {author} {\bibinfo {author} {\bibfnamefont {F.~N.}\ \bibnamefont {\"Unal}}, \bibinfo {author} {\bibfnamefont {A.}~\bibnamefont {Eckardt}},\ and\ \bibinfo {author} {\bibfnamefont {R.-J.}\ \bibnamefont {Slager}},\ }\bibfield  {title} {\bibinfo {title} {Hopf characterization of two-dimensional {F}loquet topological insulators},\ }\href {https://doi.org/10.1103/PhysRevResearch.1.022003} {\bibfield  {journal} {\bibinfo  {journal} {Phys. Rev. Res.}\ }\textbf {\bibinfo {volume} {1}},\ \bibinfo {pages} {022003(R)} (\bibinfo {year} {2019})}\BibitemShut {NoStop}%
\end{thebibliography}%

\end{document}